\documentclass[conference]{IEEEtran}

\usepackage{tikz}
\usetikzlibrary{calc}
\usetikzlibrary{arrows.meta}
\usetikzlibrary{decorations.pathreplacing}
\usetikzlibrary{fit}
\usetikzlibrary{positioning}

\usepackage{amsmath}
\usepackage{pifont}
\usepackage{bm}
\usepackage[capitalize]{cleveref}
\usepackage{algorithm}
\usepackage[noend]{algpseudocode}
\usepackage{ifthen}
\usepackage{comment}
\usepackage{subcaption}
\usepackage{bbding}
\usepackage{amsfonts}
\usepackage{anyfontsize}
\usepackage{authblk}

\newcommand{\name}{ALBUS}
\newcommand{\acronym}{Adaptive Leaky-Bucket Undulation Sensor}

\newif\ifremovecomments
\removecommentsfalse

\makeatletter
\newcommand{\mysubscript}[2]{\textsubscript{\textcolor{#2}{\textsf{\textbf{#1}}}}}
\newcommand*\defcomment[4]{
  \ifremovecomments
    \expandafter\newcommand\csname #1\endcsname[1]{%
    }
    \expandafter\newcommand\csname @#2delnoname\endcsname[1]{%
    }
    \expandafter\newcommand\csname #2del\endcsname[1]{%
    }
    \expandafter\newcommand\csname #2sugg\endcsname[1]{##1}
    \expandafter\newcommand\csname #2subs\endcsname[2]{##2}
  \else
    \expandafter\newcommand\csname #1\endcsname[1]{%
      \textcolor{#4}{\ding{110}\mysubscript{#3}{#4}\,{##1}}%
    }
    \expandafter\newcommand\csname @#2delnoname\endcsname[1]{%
      \bgroup\markoverwith{\textcolor{#4}{\rule[0.35ex]{2pt}{1pt}}}\ULon{##1}%
    }
    \expandafter\newcommand\csname #2del\endcsname[1]{%
      \csname @#2delnoname\endcsname{##1}\kern0.1em\mysubscript{#3}{#4}%
    }
    \expandafter\newcommand\csname #2sugg\endcsname[1]{%
      \textcolor{#4}{[##1]\mysubscript{#3}{#4}}%
    }
    \expandafter\newcommand\csname #2subs\endcsname[2]{%
      \csname @#2delnoname\endcsname{##1}\csname #2sugg\endcsname{##2}%
    }
  \fi
  \expandafter\newcommand\csname #2sout\endcsname{\csname #2del\endcsname}
}
\makeatother
\defcomment{simon}{ss}{SS}{magenta}
\defcomment{adrian}{ap}{AP}{blue}
\defcomment{hc}{hc}{HC}{red}

\newcommand*\circled[1]{\tikz[baseline=(char.base)]{
            \node[shape=circle,draw,inner sep=1pt] (char) {#1};}}

\newboolean{fullpaper}
\setboolean{fullpaper}{true}

\begin{document}

\date{}

\title{\name: a Probabilistic Monitoring Algorithm\\to Counter Burst-Flood Attacks}

\makeatletter
\renewcommand\AB@affilsepx{\hspace{10pt} \protect\Affilfont}
\makeatother

\author[1]{Simon Scherrer}
\author[2]{Jo Vliegen}
\author[2]{Arish Sateesan}
\author[3]{Hsu-Chun Hsiao}
\author[3,4]{Nele Mentens}
\author[1]{Adrian Perrig\vspace{-8pt}}
\affil[1]{ETH Zurich}
\affil[2]{KU Leuven}
\affil[3]{National Taiwan University}
\affil[4]{Leiden University}

\maketitle
\thispagestyle{plain}
\pagestyle{plain}

\begin{abstract}

Modern DDoS defense systems rely on probabilistic monitoring algorithms to identify flows that exceed a volume threshold and should thus be penalized. Commonly, classic sketch algorithms are considered sufficiently accurate for usage in DDoS defense. However, as we show in this paper, these algorithms achieve poor detection accuracy under burst-flood attacks, i.e., volumetric DDoS attacks composed of a swarm of medium-rate sub-second traffic bursts. Under this challenging attack pattern, traditional sketch algorithms can only detect a high share of the attack bursts by incurring a large number of false positives. 

In this paper, we present ALBUS, a probabilistic monitoring algorithm that overcomes the inherent limitations of previous schemes: ALBUS is highly effective at detecting large bursts while reporting no legitimate flows, and therefore improves on prior work regarding both recall and precision. Besides improving accuracy, ALBUS scales to high traffic rates, which we demonstrate with an FPGA implementation, and is suitable for programmable switches, which we showcase with a P4 implementation.
    
\end{abstract}

\section{Introduction}
\label{sec:intro}

As distributed denial-of-service (DDoS) attacks
continue to plague today's Internet infrastructure,
recent research has produced a range of
powerful DDoS defense systems.
The most prominent examples of such systems
include Poseidon~\cite{zhang2020poseidon},
Ripple~\cite{xing2021ripple}, Jaqen~\cite{liu2021jaqen}, COLIBRI~\cite{giuliari2021colibri}, and
ACC-Turbo~\cite{alcoz2022aggregate}.
Such DDoS defense systems provide
data-plane functionality for
\emph{detection} of attack traffic and \emph{mitigation} of the attack,
where the defense specifics
are configurable by the network operator.
In the detection component, probabilistic
monitoring algorithms (mostly the CountMin-Sketch~\cite{Cormode2005} and
the CountSketch~\cite{charikar2002finding})
monitor flows within limited memory, 
derive approximate flow-size estimates,
and report flows that violate
a volume threshold set by the network operator.
In the mitigation component, the suspicious flows 
are then blocked, rate-limited,
or deprioritized.

\begin{figure}[t]
    \centering
    \includegraphics[width=\linewidth,trim={0 10 0 0}]{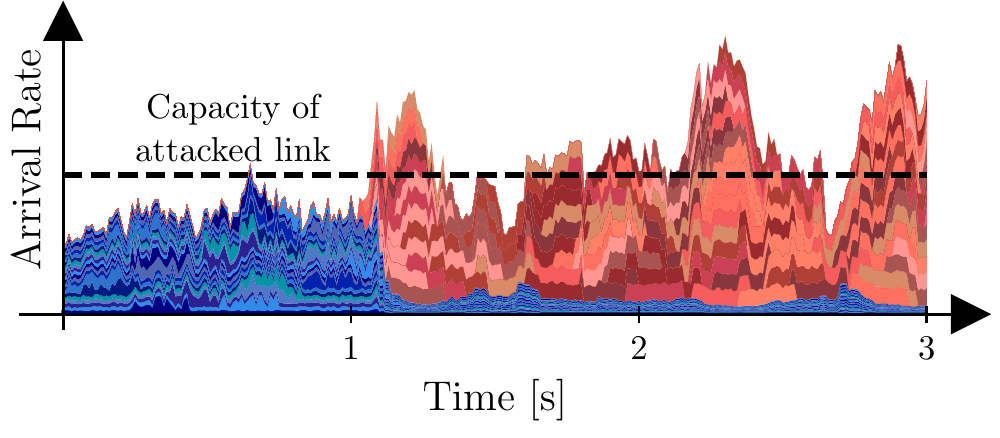}
    \caption{Burst-flood attack:
    Multiple attacker flows (in red) send simultaneous bursts. The legitimate
    flows (in blue) experience packet loss and reduce their sending rate.}
    \label{fig:introduction:attack-illustration}
    \vspace{-10pt}
\end{figure}

To provide effective defense,
state-of-the-art DDoS defense systems
rely on the accuracy of the built-in
monitoring algorithms. 
This accuracy 
is essential to construct
mitigations that are both 
comprehensive (restrict all attack traffic) and 
targeted (minimize the impact on benign traffic).
However, we identify a class of attack patterns that 
disrupt the accuracy of the monitoring
algorithms currently used in DDoS defense. 
In particular, we introduce \emph{burst-flood attacks},
i.e., volumetric DDoS attacks composed of numerous 
simultaneous bursts, where each burst is sent
in a different flow, lasts 
a few hundred milliseconds, and is only 
marginally larger than the natural bursts
of benign flows 
(cf.~\cref{fig:introduction:attack-illustration}).
We demonstrate
that these attacks lead to an 
ugly trade-off
when configuring a threshold
for common monitoring algorithms:
Detecting a large share of attacker
bursts comes at the cost of reporting
flows that do not actually violate
the allowance. 
In fact, the false reporting
of flows can only be eliminated
if almost no bursts are reported at all.
This poor performance is linked to the
regular resets which the sketch algorithms
have to perform, as these resets conflict
with the arbitrary timing and duration
of bursts.

Based on these insights, we 
develop~\name{} (\acronym), a probabilistic
monitoring algorithm that substantially
improves detection accuracy
under burst-flood attacks.
Similar to previous algorithms,
\name{} monitors all flows
in shared counters to limit memory consumption.
Crucially, however, \name{} does not
directly derive flow-volume estimates from
these counters (as sketch algorithms do),
but only leverages these counters
to continuously select the flows that are individually monitored
by exact counters.
As a result, \name{}
never reports flows 
that do not violate the configured volume allowance.
At the same time, \name{} consistently
detects a high share of excessive bursts
by avoiding resets and
applying filtering techniques.

Thanks to its design,~\name{} is well-suited
for integration into DDoS defense systems, which
only allocate limited memory to their monitoring
primitives and avoid expensive per-packet
processing.
We demonstrate \name{}'s suitability 
in three respects. First, we conduct an
extensive accuracy evaluation
to demonstrate that
\name{} generally outperforms
previous monitoring algorithms
in the scenario of interest, 
given limited memory.
Second, we also confirm the
processing efficiency of \name{} with
an implementation for a Xilinx FPGA,
with a processing capacity of 200 million packets
per second. Third, since many DDoS defense
systems are implemented in P4 and run
on programmable 
switches~\cite{zhang2020poseidon,xing2021ripple,liu2021jaqen,alcoz2022aggregate}, we also
provide a P4 implementation to support the 
integration of \name{} 
into existing systems.

In summary, this paper makes the following contributions:

\textbf{Problem description.}
    We characterize burst-flood attacks,
    demonstrate that commonly used 
    monitoring algorithms provide
    insufficient defense against such attacks,
    and elicit requirements
    that a more effective algorithm
    must fulfill.

\textbf{Algorithm design and evaluation.}
    We design the \name{} algorithm in line
    with the requirements of burst detection,
    and analyze the algorithm w.r.t.
    accuracy, security, and complexity.
    Our experiments confirm the detection
    accuracy, the attack robustness, and the
    processing efficiency.

\textbf{Implementation.}
    We implement \name{} on a Xilinx FPGA,
    thereby demonstrating that the algorithm
    is suitable for high-speed packet processing,
    and in P4, thereby confirming the straightforward
    implementation of \name{} on modern programmable switches.
\section{Problem Statement}
\label{sec:problem}

\subsection{Background}
\label{sec:problem:background}

\textbf{Burst-flood attacks.} 
Volumetric DDoS attacks attempt to exhaust network resources
(e.g., links or servers) through the sheer number of requests
from multiple attacker nodes,
thereby denying service to users of the targeted resource.
The exact structure of attacks can vary in three main ways.
First, attacks differ in the network protocol leveraged for the attack,
i.e., attack packets take the form various network protocols, 
e.g., NTP, DNS, ICMP, or bare UDP~\cite{sharafaldin2019developing}.
Second, attacks differ in the timing of aggregate attack traffic, 
e.g., pulse-wave DDoS attacks concentrate the attack traffic
in bursts lasting a few seconds~\cite{ddosguard2019pulsewave,zeifman2017attackers,alcoz2022aggregate}.
Third, attacks differ in the distribution of attack traffic
across \emph{flows}, i.e., in how the individual attacker nodes 
are scheduling and addressing their requests to create
the desired aggregate attack traffic.

In this paper, we focus on the third aspect.
More precisely, we remain agnostic regarding the content of the attack packets
and the shape of the aggregate attack traffic.
Instead, we are interested in how attackers can allot
the attack traffic to flows in order to evade modern DDoS defense systems.
Clearly, attacks using a low number of
high-rate long-lived flows 
may likely be detected rapidly. 
Therefore, most DDoS attacks today
rely on a large number of medium-rate short-lived flows, frequently
created by reflection techniques~\cite{rasti2015temporal,rossow2014amplification,griffioen2021scan}. 
For brevity, we refer to attacks with such a flow-size distribution
as \emph{burst-flood attacks} in this paper.

\textbf{DDoS defense systems.} Recent years have seen a series of proposals for DDoS defense systems, 
i.e., data-plane applications that try to maintain connectivity 
for legitimate traffic under DDoS attacks.
Since such applications require high forwarding capacity 
(to process arriving traffic at line rate) and
flexibility (to target the defense at the ongoing attack),
modern programmable switches (e.g., Tofino~\cite{agrawal2020intel}) 
have been a key enabler for most DDoS defense systems.
In particular, Poseidon~\cite{zhang2020poseidon},
Ripple~\cite{xing2021ripple}, Jaqen~\cite{liu2021jaqen} and
ACC-Turbo~\cite{alcoz2022aggregate} run on programmable switches
to both detect and mitigate DDoS attacks.
In the detection component, these DDoS defense systems try to
identify the flows that are responsible for the excessive load,
namely by subjecting flows (or other meaningful aggregates)
to monitoring algorithms. In the mitigation component,
the forwarding is adapted in order to block, rate-limit,
or deprioritize the suspicious flows.
Another approach is followed by COLIBRI~\cite{giuliari2021colibri},
which is based on source authentication and flow-based reservations.
Whenever a flow overuses its reservation, 
the flow is blocked.

Despite their differences, all these schemes refrain from
deterministic per-flow monitoring algorithms (such as NetFlow~\cite{Netflow})
to respect fast-memory limits.
Instead, all schemes rely on \emph{probabilistic} monitoring algorithms, i.e.,
algorithms that use limited memory, but only
provide a rough estimate of the traffic volume associated with a flow.
In particular, the classic CountMin-Sketch~\cite{Cormode2005} and
CountSketch~\cite{charikar2002finding} are widely used.
In this paper, we investigate the effectiveness of these
monitoring algorithms under burst-flood attacks, and
find that these algorithms are vulnerable to such attacks (cf.~\cref{sec:problem:motivation}).

\subsection{Problem Definition}
\label{sec:problem:definition}

\textbf{Threat model.}
Fundamentally, we consider an adversary that tries to exhaust
the capacity of a network resource with
high-rate network traffic.
We formalize traffic to be composed
of \emph{packets}, 
where each packet~$p$ arrives
at a certain time~$t_p$, has a certain size~$s_p$, and 
can be assigned to a \emph{flow}~$f_p$ 
based on properties from the packet header.
As a result, the traffic volume sent by flow~$f$
in a time window~$[t_1, t_2]$ is:
\begin{equation}
    \sigma(f, t_1, t_2) =  \sum\nolimits_{p \text{ s.t. } f_p = f\ \land\ t_p \in [t_1, t_2]} s_p.
\end{equation}
In our scenario, the packets of each attack flow
are concentrated in time, i.e.,
form a \emph{burst}.

To formalize bursts, we note that a burst is commonly understood
as a temporary increase of the sending rate compared
to a base rate~$\gamma$. Hence, if a flow~$f$
sends traffic volume~$\sigma(f, t_1, t_2) > \gamma (t_2 - t_1)$ 
during a time window~$[t_1, t_2]$, 
then we consider~$b = \sigma(f, t_1, t_2) - \gamma (t_2 - t_1)$ the burst size.
A threshold on the burst size~$b$ is thus always given
in the form of a \emph{flow specification}~$\gamma t + \beta$, 
where~$\beta \geq 0$ is the burstiness allowance. 
Concretely, if a flow sends more traffic than~$\gamma (t_2 - t_1) + \beta$ 
during a time window~$[t_1, t_2]$, 
we consider it \emph{excessively bursty}.
This formalization is the classic burstiness definition 
in networking, reflects queuing dynamics, and is independent
of fixed time windows (unlike other burstiness definitions~\cite{zhong2021burstsketch}).
Moreover, it allows to distinguish excessive burstiness
from moderate burstiness; the latter is a feature
of almost all Internet traffic.

We extend this definition for our purpose:
A burst of 
\emph{width~$w$} and \emph{overuse ratio~$\ell$}
is a packet sequence of a single flow~$f$,
spanning a time window of length~$w$ and
containing traffic volume~$\gamma w + \ell \beta$, i.e.,
$\sigma(f, t, t + w) = \gamma w + \ell\beta$.
The adversary can create
bursts of arbitrary size and duration.

\textbf{Objective.} Our goal is to design a monitoring algorithm that
reports a set~$B_r$ of bursts,
which ideally should match the set~$B$ 
of all excessive bursts:
\begin{equation}
    B = \left\{(f, t_1, t_2)\ \mid\ \sigma(f, t_1, t_2) > \gamma(t_2 - t_1) + \beta\right\}.
\end{equation}
Hence, our algorithm can be used
by DDoS defense systems, which
configure monitoring primitives with a threshold.

\textbf{Metrics.} We measure reporting
accuracy with two conventional metrics.
First, \emph{recall} is
the share of allowance-violating
bursts that are reported, i.e., 
$\mathit{recall} = |B_r \cap B|/|B|$.
If a burst exceeds the allowance,
but is not reported, it constitutes
a \emph{false negative} in our setting ($B \setminus B_r$).
Second, \emph{precision} is the share of allowance-violating
bursts among all reported bursts, i.e.,
$\mathit{precision} = |B_r \cap B|/|B_r|$.
If a burst does not exceed the allowance,
but is reported, it constitutes
a \emph{false positive} in our context ($B_r \setminus B$).
Precision and recall can also be combined
into the $F_1$ score, which is 1 only if~$B = B_r$,
i.e., $F_1 = 2 |B_r \cap B| / (|B_r| + |B|)$.

\subsection{Targeted Deployment Setting}

We design an algorithm to be included
as a monitoring primitive in
existing DDoS defense systems.
These DDoS defense systems provide
the following relevant functionality:

\textbf{Flow definition.} The exact definition of a flow is at the discretion
of the DDoS defense system using our monitoring algorithm, 
and may depend on the ongoing attack.
For example, the DDoS defense system might consider packets
belonging to the same flow if they share the destination IP prefix (e.g., in UDP carpet bombing)
or the full fivetuple (e.g., in UDP flood)~\cite{alcoz2022aggregate}.
The monitoring algorithm simply accepts the flow identifier
that the incoming packets are tagged with.

\textbf{Threshold definition.} 
The monitoring algorithm is configured to
report flows that exceed a \emph{volume threshold}.
This threshold should be defined such that the algorithm
reports the exact set of excessively bursty flows.
\name{} accepts thresholds of the form~$(\gamma,\beta)$
directly matching the allowance, whereas threshold definition
for previous algorithms is more involved (cf.~\S\ref{sec:problem:motivation}).
Hence, the notion of accuracy in this paper
is based on \emph{volume}, i.e.,
accuracy means reporting the \emph{allowance-violating} flows.
Accuracy in terms of \emph{intention}, 
i.e., reporting \emph{malicious} flows,
is thus achieved with an appropriate threshold
if malicious flows are identifiable by volume.
DDoS defense systems may find such a threshold
in various ways:

\textit{Signatures.} Many DDoS defense systems~\cite{zhang2020poseidon,xing2021ripple,liu2021jaqen} 
are based on descriptions of known attacks (signatures),
and compare them to current observations.
The threshold may be chosen such that the attack flows
in the signature violate it.

\textit{Calibration with queries.} Jaqen~\cite{liu2021jaqen} allows 
the network operator to submit \emph{queries}, and
reports all flows in the current traffic that match
the query, e.g., exceed a volume threshold.
For \name{}, a network operator may perform the following query-based
calibration (when not under attack) to find suitable~$(\gamma,\beta)$:
The operator can choose a base rate~$\gamma$, submit a series
of queries with increasing burstiness allowance~$\beta$,
and observe the decreasing share of reported flows.
Then, the threshold~$(\gamma,\beta)$ is appropriate when
the number of benign flows violating it has been reduced
to an acceptable amount.

\textit{Reservations.} COLIBRI~\cite{giuliari2021colibri}
maintains bandwidth reservations for a guaranteed data rate
for individual flows. This data rate then constitutes 
the allowed base rate~$\gamma$.
The burstiness allowance~$\beta$ is $\beta = (u-1)\gamma w$ 
if flows with excessive rate~$u\gamma$, $u > 1$,
are tolerated for at most~$w$ seconds.

\begin{figure*}
\newcommand{\evallinewidthintro}{0.19}
    \begin{subfigure}{0.38\linewidth}
        \includegraphics[width=0.49\linewidth]{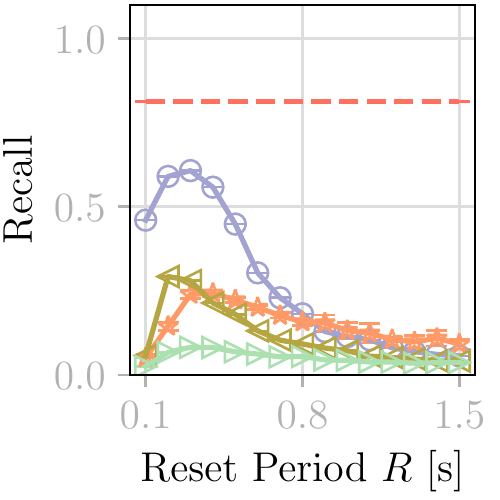}
        \includegraphics[width=0.49\linewidth]{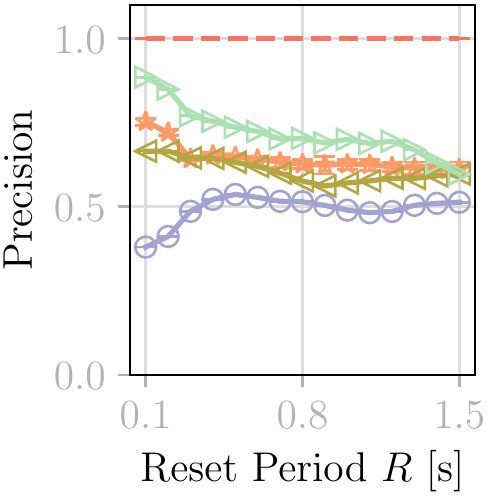}
        \caption{CountSketch.}
        \label{fig:motivation:200:cs}
    \end{subfigure}\vrule\ 
    \begin{subfigure}{0.57\linewidth}
        \includegraphics[width=0.328\linewidth]{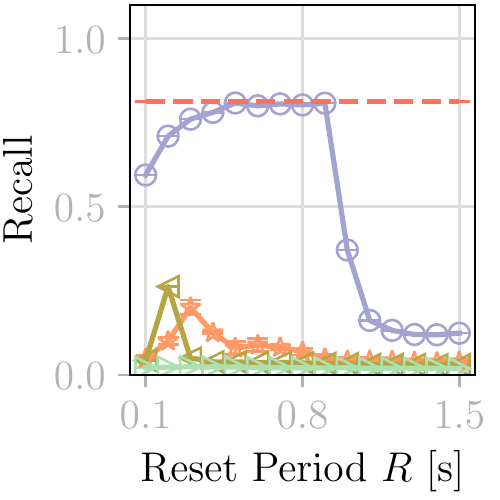}
        \includegraphics[width=0.656\linewidth]{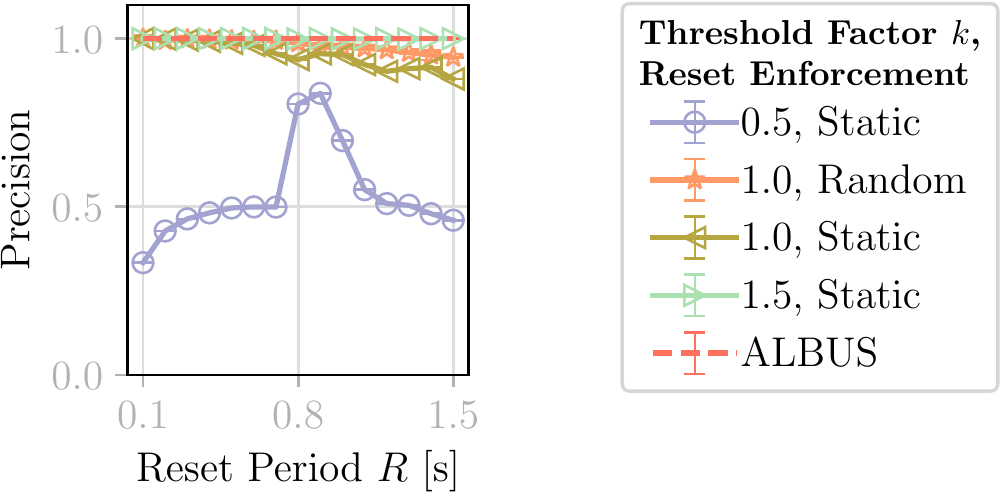}
        \caption{CountMin-Sketch.}
        \label{fig:motivation:200:cm}
    \end{subfigure}
    \vspace{-5pt}
    \centering
    \caption{Detection performance under burst-flood attack (200ms bursts).}
    \vspace{10pt}
    \label{fig:motivation:200}
\end{figure*}

\begin{figure*}
    \vspace{-15pt}
    \centering
     \begin{subfigure}{0.38\linewidth}
        \includegraphics[width=0.49\linewidth]{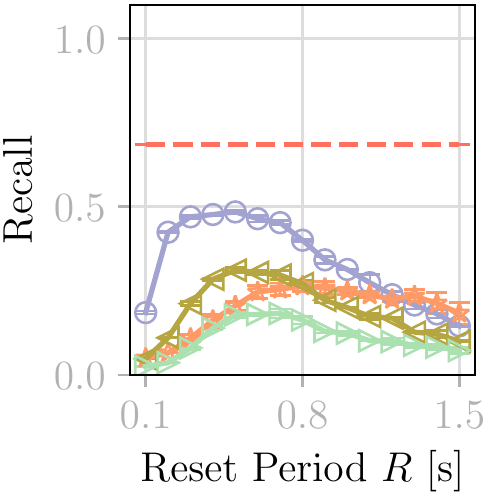}
        \includegraphics[width=0.49\linewidth]{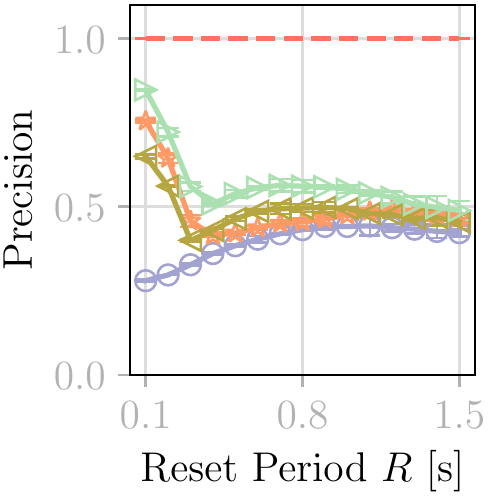}
        \caption{CountSketch.}
        \label{fig:motivation:500:cs}
    \end{subfigure}\vrule\ 
    \begin{subfigure}{0.57\linewidth}
        \includegraphics[width=0.328\linewidth]{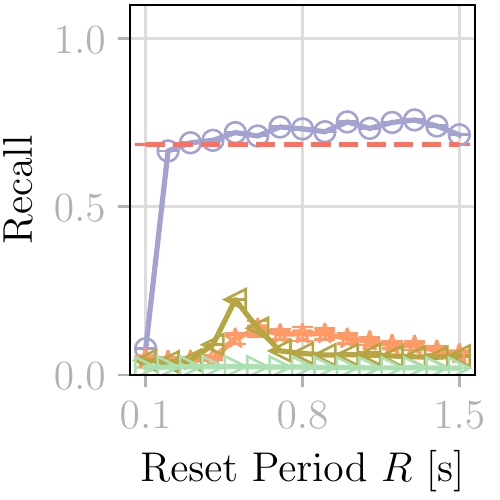}
        \includegraphics[width=0.328\linewidth]{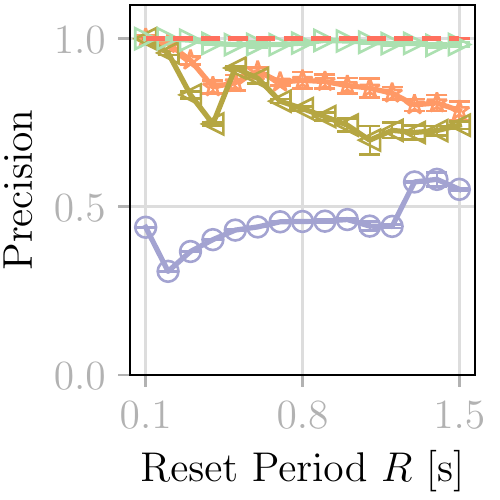}
        \includegraphics[width=0.328\linewidth]{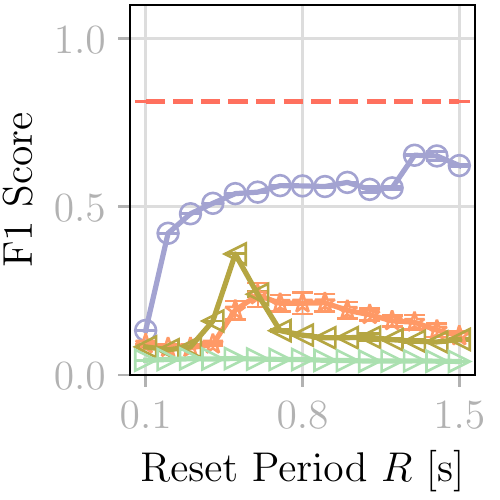}
        \caption{CountMin-Sketch.}
        \label{fig:motivation:500:cm}
    \end{subfigure}
    \vspace{-5pt}
    \caption{Detection performance under burst-flood attack (500ms bursts).}
    \label{fig:motivation:500}
\end{figure*}

\subsection{Algorithm Requirements}
\label{sec:problem:requirements}

Given the objective and 
intended usage of 
our algorithm
as well as the shortcomings of 
previous algorithms, we identify
the following algorithm requirements.

\textbf{Processing efficiency.} A DDoS attacker
can maximize the attack damage by targeting 
network resources of systemic relevance, 
most importantly routers in the Internet core.
Since such routers 
process billions of packets per second and handle
millions of flows concurrently~\cite{caida201810}, 
any monitoring algorithm
on these routers must conform to narrow 
constraints on the additional processing complexity. 
Since the required
processing efficiency does not allow main-memory 
look-ups, 
processing efficiency also limits admissible memory 
consumption to the capacity of fast cache or SRAM memory.
Hence, individual monitoring of every flow by dedicated 
counters (as in NetFlow~\cite{Netflow}),
or approaches that keep time-series
data for each flow~\cite{chen2006collaborative,fu2021realtime,luo2005new,chang2010taming}
are impractical in the network core.
Hence, DDoS defense systems use 
probabilistic flow-monitoring algorithms 
with acceptable memory consumption
and low operational overhead per packet.

\textbf{Accuracy.} 
Despite limited memory, the monitoring
algorithm should accurately identify
bursts that violate the configured
flow specification.
In particular, the algorithm should
achieve high \emph{recall} and high
\emph{precision} simultaneously (cf.~\cref{sec:problem:definition}).
A high score in both metrics allows
the DDoS defense system to
identify and eradicate attack traffic,
while limiting punitive actions to flows
that actually violate the flow specification.
For burst-flood attacks, high accuracy
is especially challenging 
to achieve, because the aggregate
attack traffic might be composed of
bursts which only marginally exceed
the burstiness allowance.

\textbf{Time-window flexibility.} 
From the experiment in~\cref{sec:problem:motivation},
we deduce that the reliance on discrete time windows
constitutes a major impediment to 
detection accuracy. Attack bursts
can be arbitrarily timed and
arbitrarily wide, which poses an issue for the numerous
monitoring algorithms that operate with a 
\emph{landmark-window model}~\cite{Estan03, sivaraman2017heavy, scherrer2021lowrate, zhong2021burstsketch, huang2021toward, sheng2021pr, Cormode2005, yang2019heavykeeper, charikar2002finding}. 
In the landmark-window model, algorithms
split time into disjoint intervals
and compare traffic measurements
to the expectations for the interval. This interval
duration is subject to an undesirable trade-off:
If the intervals are too long, even
a high-volume burst might not 
exceed the traffic volume
allowed for the whole interval if followed
by a sending pause;
if the measurement intervals are too short,
accuracy suffers as 
statistical noise grows stronger. Furthermore,
bursts could be timed to arrive at the
transition between two intervals, where
measurements are reset.
Hence, we aim to design an algorithm
without regular resets.

\section{Why Another Monitoring Algorithm?}
\label{sec:problem:motivation}

In this section, we evaluate the monitoring algorithms
present in current DDoS defense systems
under burst-flood attacks. 
In particular, we 
conduct a simulation-based experiment to
evaluate CountMin-Sketch~\cite{Cormode2005},
which is a component of Poseidon~\cite{zhang2020poseidon} and
Ripple~\cite{xing2021ripple}, and CountSketch~\cite{charikar2002finding},
which is a component of Jaqen~\cite{liu2021jaqen} 
(through UnivMon~\cite{liu2016one}).

\textbf{Experiment.}
We replay 5 seconds of a CAIDA trace and augment it
with a burst flood amouting to 5 Gbps at any point in time.
The attack bursts last $w= 200$~ms or $w = 500$~ms, and 
have an overuse ratio of~$\ell = 1.2$
(Experiment details in~\cref{sec:evaluation:setup}).
The goal of the algorithms (each obtaining 300~KB of memory) 
is to find all bursts
that violate the flow specification of~$\gamma = 1$~Mbps
and~$\beta = 50$~KB, both in the background traffic
and the synthesized attack traffic.
We measure both recall (share of detected bursts)
and precision (share of correctly detected bursts among
all reported bursts).

\textbf{CountMin-Sketch.}
The CountMin-Sketch relies on a datastructure 
containing $D$~counter arrays, where each array~$c_i$ 
is associated with a distinct hash
function~$h_i$. Upon arrival of a packet of flow~$f$,
the counter~$h_i(f)$ is increased by the packet size
in every counter array~$c_i$. The estimate for
a flow volume is then~$v_f = \min_{i \in\{1, ..., D\}} c_i[h_i(f)]$.
If~$v_f$ exceeds a configured threshold~$T$, the flow is reported.
In the following experiment, we apply
the CountMin-Sketch using \emph{reset periods}, 
i.e., time intervals 
after which the data structure is reset.
These reset periods are needed to regularly
discard old information about past bursts.
Given a reset period~$R$, we consider
\emph{static} reset enforcement (used in current DDoS defense systems), 
where the time~$R'$ between resets always equals~$R$,
and \emph{randomized} reset enforcement,
where the time~$R'$ between resets is sampled
uniformly at random from~$(0, R]$ after every reset.
We try to achieve
the detection goal by varying the threshold~$T$ 
based on the current reset timer~$R'$ and a threshold factor~$k$: 
$T = k \cdot \left(\gamma \cdot R' + \beta\right)$.
Intuitively, the CountMin-Sketch would
identify the attack bursts
for~$R' = w$~ms and~$k = 1$ if the bursts
were perfectly aligned with the reset periods;
the threshold factor is used to account
for temporal misalignment.

\textbf{CountSketch.} The CountSketch is an extension of the CountMin-Sketch.
Specifically, the CountSketch contains a second hash function~$s_i$ for
each counter array~$c_i$, where~$s_i$ maps the flow ID~$f$ to $-1$ or $+1$.
The counter~$h_i(f)$ in array~$c_i$ is then increased or decreased 
by the packet size depending on the value of~$s_i$.
The estimate for the volume of flow~$f$ is~$v_f = \mathrm{median}_{i \in\{1, ..., D\}}\ s_i(f) \cdot c_i[h_i(f)]$.
The threshold of the CountSketch is varied analogously to the
CountMin-Sketch.

\textbf{Results.} 
\Cref{fig:motivation:200,fig:motivation:500} show the experiment results 
for 200~ms and 500~ms bursts, respectively.
From these figures, we make a number of observations.
First, threshold factors that lead to good performance
on recall lead to poor performance on precision, and vice versa. 
For example, a low threshold factor
achieves high recall, but low precision.
This trade-off suggests that the sketches, based on
discrete time windows, fail to identify excessively
bursty flows in a targeted manner.
Second, precision and recall are unpredictably
related to the reset period. 
Moreover, the optimal reset period in terms of
a given metric varies between burst widths.
Given that burst-flood attacks contain bursts
of unknown and varying size, choosing an appropriate
threshold is daunting.
Third, the limitations of the sketches cannot be
overcome by simply randomizing the duration between resets,
as such randomization does not yield a clear improvement.
Fourth, we have added the performance of our algorithm, \name{}
(\name{} does not have
reset periods). This addition demonstrates
that alternative designs to sketches are promising;
a detailed comparative evaluation of~\name{} is presented
in~\cref{sec:evaluation}. 
\section{Design}
\label{sec:design}

In this section, we present \name{} (\acronym), which is
our approach to detect excessively bursty flows under the
requirements from~\cref{sec:problem:requirements}.
As \name{} is partially based on the leaky-bucket algorithm, we describe this algorithm in~\cref{sec:design:leaky-bucket}.
In~\cref{sec:design:idea}, we discuss the high-level
ideas of \name{}, and provide a detailed description of the algorithm 
in~\cref{sec:design:algorithm}.

\subsection{Background: Leaky Buckets}
\label{sec:design:leaky-bucket}

The leaky-bucket (LB) algorithm is a classic approach to
rate control in networks, with a history of usage
reaching back to ATM networks~\cite{niestegge1990leaky}.
We revisit this algorithm for its properties 
which are perfectly in line with the special requirements
of burst detection. In particular,
the LB algorithm detects any violation
of a flow specification of the 
from~$\gamma t + \beta$ (cf.~\cref{sec:problem:definition}),
i.e., reports a flow if and only if
the flow sends more traffic than~$\gamma w + \beta$
during any time window with duration~$w$.
Hence, the LB algorithm by design fulfills all of the requirements from~\cref{sec:problem:requirements},
as it is perfectly accurate and
flexible with respect to time windows.

Each LB~$\lambda$ contains a count~$\lambda.c$ and
a timestamp~$\lambda.t$. 
The LB algorithm
is akin to a physical leaky bucket, which is filled
at a varying rate, leaks at a
constant rate (unless it is empty), and overflows
if the filling rate exceeds the draining rate
for too long (cf.~\cref{alg:leaky-bucket}).
Note that the \emph{draining volume}~$d$
is of special importance to~\name{}.

\begin{algorithm}[t]
\caption{Leaky-bucket algorithm}\label{alg:leaky-bucket}
\begin{algorithmic}
\Function{Update}{Bucket $\lambda$, packet size~$s$, timestamp $t$}

\State $d \gets \gamma \cdot (t - \lambda.t)$
\State $\lambda.c \gets \max(\lambda.c - d, 0) + s$ \hspace{30pt} $\lambda.t \gets t$
\If{$\lambda.c > \beta$} \Return \text{True} 
\Else{} \Return \text{False}
\EndIf
\EndFunction
\end{algorithmic}
\end{algorithm}

\subsection{Design Idea of \name{}}
\label{sec:design:idea}

On a high level, \name{} applies
the LB algorithm to find excessive bursts. 
Given unlimited 
memory, each individual flow could be monitored 
with  a dedicated LB. To respect 
memory constraints, however,
we choose to monitor only a subset
of flows at any point in time.
The challenge thus becomes how
to dynamically select the subset
of flows individually
monitored by dedicated LBs.
We observe that this subset 
selection can be performed on a salient
property of bursty flows,
namely their \emph{consistent burstiness}:
As bursty flows send a
large flow volume in a short time,
all of their consecutive packets 
are sent 
in a bursty manner, i.e., with a rate
above allowed rate~$\gamma$.
In contrast, the moderate burstiness 
of legitimate flows rules out 
that the allowed rate is persistently
exceeded. Instead, moderate bursts are 
followed by under-utilization of the 
allowance~$\gamma$ after only a few packets.

\begin{figure}[t]
    \centering
    \begin{tikzpicture}[
flowsquare/.style={draw=black!60, shading=radial,outer color={rgb,255:red,137;green,207;blue,240},inner color=white, thick, minimum size=\nodewidth,outer sep=0pt},
flowsizesquare/.style={draw=black!60, inner color=white, thick, minimum size=\nodewidth},
indexsquare/.style={draw=black!60,  shading=radial,outer color={rgb,1:red,0.8;green,0.8;blue,0.8},inner color=white, thick,minimum width=\counterwidth,minimum height=\indexheight,align=center},
lbsquare/.style={draw=black!60, shading=radial,outer color={rgb,255:red,240;green,207;blue,137},inner color=white, thick,minimum width=\counterwidth,minimum height=\lbheight,align=center},
pcsquare/.style={draw=black!60, shading=radial,outer color={rgb,1:red,0.53;green,0.66;blue,0.42},inner color=white, thick,minimum width=\counterwidth,minimum height=\lbheight,align=center},
]

    \def\nodewidth{6mm};
    \def\counterwidth{6mm};
    \def\indexheight{5mm};
    \def\lbheight{6mm};
    
    \draw (0.5, -1.2) -- (0.5,1.9);
    \draw (3.2,-1.2) -- (3.2,1.9);
    \node (pkt_desc) at (0, -0.95) {\textbf{Pkt}};
    \node (mem_desc) at (1.85, -1.0) {\textbf{Memory}};
    \node (mem_desc) at (5.5, -1.0) {\textbf{Decision logic}};
    
    \node[flowsquare] (f1) at (0,0.5) {$f_1$};
    
    \node (lb_desc) at (1.9,1.7) {Leaky buckets $\Lambda$};
    \foreach \x in {0,...,3} {
        \pgfmathsetmacro\xpos{0.95+\x*\counterwidth/2.83465/10}
        \node[lbsquare] at (\xpos,1) (l\x) {};
    };
    
    \node at (l1.center) {$f_1$};
    
    \node[align=center] (pc_desc) at (1.85,-0.5) {\fontsize{7.8pt}{\baselineskip}\selectfont Background counters $\Pi$};
    \foreach \x in {0,...,3} {
        \pgfmathsetmacro\xpos{0.95+\x*\counterwidth/2.83465/10}
        \node[pcsquare] at (\xpos,0) (p\x) {};
    };
    
    \node at (p1.center) {$f_2$};
    
    \draw[-latex] (f1.east) -- (l1.west);
    
    \node[indexsquare,align=center,rounded corners=0.2cm] (cond) at (4.2,1) {Net inflow\\positive?};
    \draw[dashed] (l1.north west) -- (cond.145);
    \draw[dashed] (l1.south west) -- (cond.215);
    
    \node[indexsquare,rounded corners=0.2cm] (yes_result) at (6.7, 1) {Update LB};
    \draw[dotted,-latex,thick] (cond.east) -- (yes_result.west);
    \node (yes) at (5.4, 1.2) {Yes};
    
    \node[indexsquare,align=center,rounded corners=0.2cm] (no_result) at (6.7, 0) {Evict~$f_1$ and\\pull from BC};
    \draw[dotted,-latex,thick] (cond.325) -- (no_result.151);
    \node (no) at (5.4, 0.75) {No};
    
    \draw[dotted,-latex,thick] (no_result) -- (p1.east);
    \draw[dotted,-latex,thick] (p1.north) -- (l1.south);

\end{tikzpicture}
    \caption{Basic mechanism of \name{}.}
    \label{fig:design:idea}
    \vspace{-10pt}
\end{figure}
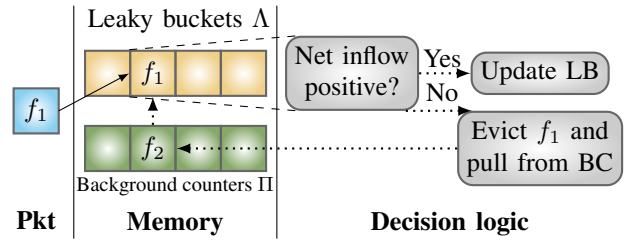

This observation inspires the 
central criterion used for adapting
the subset of LB-monitored flows
(cf.~\cref{fig:design:idea}). 
Given a set~$\Lambda$
of LBs, a flow is deterministically
mapped to an LB~$\lambda$ by a hash on the 
flow ID~$f$. If this LB~$\lambda$ is not assigned
to any other flow, it can operate as a dedicated
LB for flow~$f$. 
Once assigned to an LB, flow~$f$ is
monitored as long as the \emph{net inflow}
(packet size minus draining volume) to
the LB is positive. As soon as
the net inflow is negative, flow~$f$ is
evicted from the LB, which can then 
monitor another flow.

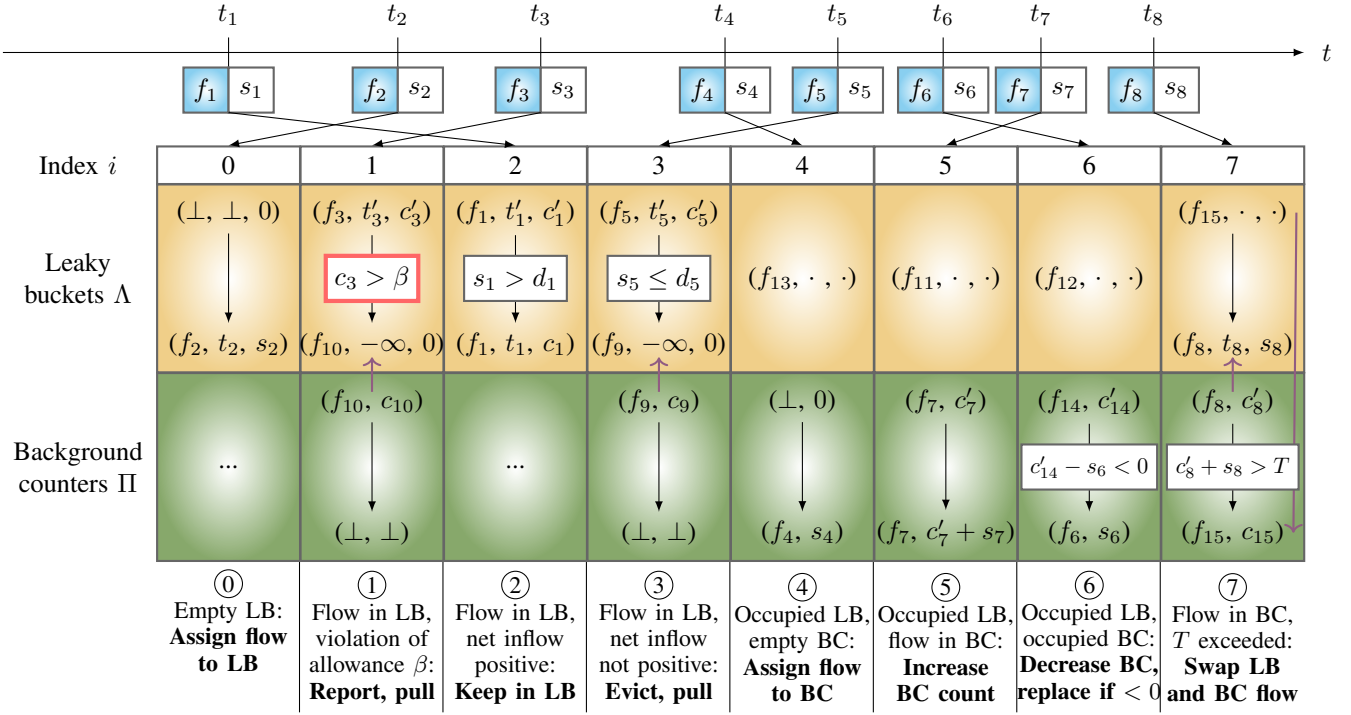
\begin{figure*}[t!]
    \centering
    \begin{tikzpicture}[
flowsquare/.style={draw=black!60, shading=radial,outer color={rgb,255:red,137;green,207;blue,240},inner color=white, thick, minimum size=\nodewidth,outer sep=0pt},
flowsizesquare/.style={draw=black!60, inner color=white, thick, minimum size=\nodewidth},
indexsquare/.style={draw=black!60, inner color=white, thick,minimum width=\counterwidth,minimum height=\indexheight,align=center},
lbsquare/.style={draw=black!60, shading=radial,outer color={rgb,255:red,240;green,207;blue,137},inner color=white, thick,minimum width=\counterwidth,minimum height=\lbheight,align=center},
pcsquare/.style={draw=black!60, shading=radial,outer color={rgb,1:red,0.53;green,0.66;blue,0.42},inner color=white, thick,minimum width=\counterwidth,minimum height=\lbheight,align=center},
]

    \definecolor{antiquefuchsia}{rgb}{0.57, 0.36, 0.51}

    \def\nodewidth{6mm};
    \def\counterwidth{19mm};
    \def\indexheight{5mm};
    \def\lbheight{25mm};

    \draw[-latex] (-2, 10) -- (15.3, 10);
    \node (arrow_text) at (15.6, 10) {$t$};
    
    \foreach \x/\xpos in {1/1, 2/3.25, 3/5.15, 4/7.6, 5/9.1, 6/10.5, 7/11.8, 8/13.3} {
        \node at (\xpos, 10) (t\x) {};
        \node at (\xpos, 10.5) (t\x_text) {$t_{\x}$};
        \draw ($(t\x)-(0,2mm)$) -- ($(t\x)+(0,2mm)$);
        \node[flowsquare] at ($(t\x.center)-(0.5*\nodewidth,5mm)$) (f\x) {$f_{\x}$};
        \node[flowsizesquare] at ($(f\x.center)+(\nodewidth,0)$) (f\x_size) {$s_{\x}$};
    };
    
    \node at (-1, 8.5) (index_desc) {Index $i$};
    \foreach \x in {0,...,7} {
        \pgfmathsetmacro\xpos{1+\x*\counterwidth/2.83465/10}
        \node[indexsquare] at (\xpos,8.5) (i\x) {\x} {};
    };
    
    \pgfmathsetmacro\centerdist{(0.5*\indexheight+0.5*\lbheight)/2.83465/10};
    \node[align=center] (lb_desc) at ($(index_desc.center) - (0,\centerdist)$) {Leaky\\buckets $\Lambda$};
    \foreach \x in {0,...,7} {
        \node[lbsquare] at ($(i\x.center)-(0,\centerdist)$) (l\x) {};
    };
    
    \draw[-latex] (f2.south east) -- (i0.north);
    \node[align=center] (l0_old) at ($(l0.center) + (0, 0.35*\lbheight)$) 
    {($\bot$, $\bot$, $0$)};
    \node[align=center] (l0_new) at ($(l0.center) - (0, 0.35*\lbheight)$) 
    {($f_2$, $t_2$, $s_2$)};
    \draw[-latex] (l0_old.south) -- (l0_new.north);
    
    \draw[-latex] (f3.south east) -- (i1.north);
    \node[align=center] (l1_old) at ($(l1.center) + (0, 0.35*\lbheight)$) 
    {($f_3$, $t_3'$, $c_3'$)};
    \node[align=center] (l1_new) at ($(l1.center) - (0, 0.35*\lbheight)$) 
    {($f_{10}$, $-\infty$, $0$)};
    \draw[-latex] (l1_old.south) -- (l1_new.north);
    \node[flowsizesquare,draw=red!60,ultra thick] (l1_calc) at (l1.center) {$c_3 > \beta$};
    
    \draw[-latex] (f1.south east) -- (i2.north);
    \node[align=center] (l2_old) at ($(l2.center) + (0, 0.35*\lbheight)$) 
    {($f_1$, $t_{1}'$, $c_1'$)};
    \node[align=center] (l2_new) at ($(l2.center) - (0, 0.35*\lbheight)$) 
    {($f_1$, $t_1$, $c_1$)};
    \draw[-latex] (l2_old.south) -- (l2_new.north);
    \node[flowsizesquare] (l2_calc) at (l2.center) {$s_1 > d_1$};
    
    \draw[-latex] (f5.south east) -- (i3.north);
    \node[align=center] (l3_old) at ($(l3.center) + (0, 0.35*\lbheight)$) 
    {($f_5$, $t_5'$, $c_5'$)};
    \node[align=center] (l3_new) at ($(l3.center) - (0, 0.35*\lbheight)$) 
    {($f_9$, $-\infty$, $0$)};
    \draw[-latex] (l3_old.south) -- (l3_new.north);
    \node[flowsizesquare] (l3_calc) at (l3.center) {$s_5 \leq d_5$};
    
    \draw[-latex] (f4.south east) -- (i4.north);
    \node[align=center] (l4_old) at (l4.center) 
    {($f_{13}$, $\cdot$ , $\cdot$)};
    
    \draw[-latex] (f7.south east) -- (i5.north);
    \node[align=center] (l5_old) at (l5.center) 
    {($f_{11}$, $\cdot$ , $\cdot$)};
    
    \draw[-latex] (f6.south east) -- (i6.north);
    \node[align=center] (l6_old) at (l6.center) 
    {($f_{12}$, $\cdot$ , $\cdot$)};
    
    \draw[-latex] (f8.south east) -- (i7.north);
    \node[align=center] (l7_old) at ($(l7.center) + (0, 0.35*\lbheight)$) 
    {($f_{15}$, $\cdot$ , $\cdot$)};
    \node[align=center] (l7_new) at ($(l7.center) - (0, 0.35*\lbheight)$) 
    {($f_8$, $t_8$, $s_8$)};
    \draw[-latex] (l7_old.south) -- (l7_new.north);

    \pgfmathsetmacro\centerdist{\lbheight/2.83465/10};
    \node[align=center] (pf_desc) at ($(lb_desc.center) - (0,\centerdist)$) {Background\\counters $\Pi$};
    \foreach \x in {0,...,7} {
        \node[pcsquare] at ($(l\x.center)-(0,\centerdist)$) (p\x) {};
    };
    
    \node (p0_empty) at (p0.center) {...};
    \node (p3_empty) at (p2.center) {...};
    
    \node[align=center] (p1_old) at ($(p1.center) + (0, 0.35*\lbheight)$) 
    {($f_{10}$, $c_{10}$)};
    \node[align=center] (p1_new) at ($(p1.center) - (0, 0.35*\lbheight)$) 
    {($\bot$, $\bot$)};
    \draw[-latex] (p1_old.south) -- (p1_new.north);
    \draw[->,color=antiquefuchsia,thick] ($(p1.center) + (0, 0.4*\lbheight)$) -- ($(l1.center) - (0, 0.42*\lbheight)$);
    

    \node[align=center] (p3_old) at ($(p3.center) + (0, 0.35*\lbheight)$) 
    {($f_{9}$, $c_{9}$)};
    \node[align=center] (p3_new) at ($(p3.center) - (0, 0.35*\lbheight)$) 
    {($\bot$, $\bot$)};
    \draw[-latex] (p3_old.south) -- (p3_new.north);
    \draw[->,color=antiquefuchsia,thick] ($(p3.center) + (0, 0.4*\lbheight)$) -- ($(l3.center) - (0, 0.42*\lbheight)$);
    
    \node[align=center] (p4_old) at ($(p4.center) + (0, 0.35*\lbheight)$) 
    {($\bot$, $0$)};    
    \node[align=center] (p4_new) at ($(p4.center) - (0, 0.35*\lbheight)$) 
    {($f_4$, $s_4$)};
    \draw[-latex] (p4_old.south) -- (p4_new.north);
    
    \node[align=center] (p5_old) at ($(p5.center) + (0, 0.35*\lbheight)$) 
    {($f_7$, $c_7'$)};
    \node[align=center] (p5_new) at ($(p5.center) - (0, 0.35*\lbheight)$) 
    {($f_7$, $c_7' + s_7$)};
    \draw[-latex] (p5_old.south) -- (p5_new.north);
    
    \node[align=center] (p6_old) at ($(p6.center) + (0, 0.35*\lbheight)$) 
    {($f_{14}$, $c_{14}'$)};
    \node[align=center] (p6_new) at ($(p6.center) - (0, 0.35*\lbheight)$) 
    {($f_6$, $s_6$)};
    \draw[-latex] (p6_old.south) -- (p6_new.north);
    \node[flowsizesquare] (p6_calc) at (p6.center) {\scalebox{0.85}{$c_{14}' - s_6 < 0$}};
    
    \node[align=center] (p7_old) at ($(p7.center) + (0, 0.35*\lbheight)$) 
    {($f_{8}$, $c_{8}'$)};
    \node[align=center] (p7_new) at ($(p7.center) - (0, 0.35*\lbheight)$) 
    {($f_{15}$, $c_{15}$)};
    \draw[-latex] (p7_old.south) -- (p7_new.north);
    \draw[->,color=antiquefuchsia,thick] (l7_old.east) -- (p7_new.east);
    \node[flowsizesquare] (p7_calc) at (p7.center) {\scalebox{0.85}{$c_{8}' + s_8 > T$}};
    \draw[->,color=antiquefuchsia,thick] ($(p7.center) + (0, 0.4*\lbheight)$) -- ($(l7.center) - (0, 0.42*\lbheight)$);
    

    \pgfmathsetmacro\centerdist{(0.5*\indexheight+0.5*\lbheight)/2.83465/10};
    \foreach \x in {1,...,7} {
        \draw[-] ($(p\x.south west)$) -- ($(p\x.south west) - (0, 0.8*\lbheight)$);
    };

    \node[align=center,font=\small] (ann0) at ($(p0.south) - (0, 0.31*\lbheight)$) {\circled{0}\\Empty LB:\\\textbf{Assign flow}\\\textbf{to LB}}; 

    \node[align=center,font=\small] (ann1) at ($(p1.south) - (0, 0.4*\lbheight)$) {\circled{1}\\Flow in LB,\\violation of\\allowance $\beta$:\\\textbf{Report, pull}}; 

    \node[align=center,font=\small] (ann2) at ($(p2.south) - (0, 0.4*\lbheight)$) {\circled{2}\\Flow in LB,\\ net inflow\\positive:\\\textbf{Keep in LB}}; 

    \node[align=center,font=\small] (ann3) at ($(p3.south) - (0, 0.4*\lbheight)$) {\circled{3}\\Flow in LB,\\ net inflow\\not positive:\\\textbf{Evict, pull}}; 

    \node[align=center,font=\small] (ann4) at ($(p4.south) - (0, 0.4*\lbheight)$) {\circled{4}\\Occupied LB,\\ empty BC:\\\textbf{Assign flow}\\\textbf{to BC}}; 

    \node[align=center,font=\small] (ann5) at ($(p5.south) - (0, 0.4*\lbheight)$) {\circled{5}\\Occupied LB,\\flow in BC:\\\textbf{Increase}\\\textbf{BC count}}; 

    \node[align=center,font=\small] (ann6) at ($(p6.south) - (0, 0.4*\lbheight)$) {\circled{6}\\Occupied LB,\\ occupied BC:\\\textbf{Decrease BC,}\\\textbf{replace if $<0$}}; 

    \node[align=center,font=\small] (ann7) at ($(p7.south) - (0, 0.4*\lbheight)$) {\circled{7}\\Flow in BC,\\$T$ exceeded:\\\textbf{Swap LB}\\\textbf{and BC flow}};

\end{tikzpicture}
    \caption{Detailed overview of case distinctions in \name{} (Discussion in~\cref{sec:design:algorithm}).}
    \vspace{-10pt}
    \label{fig:design:algorithm}
\end{figure*}

If reduced to this central mechanism,
\name{} would suffer from an issue
in terms
of both security and accuracy.
In particular, even an extremely
bursty flow would never be detected in case
another flow already occupies the LB 
to which the bursty flow is mapped.
An attacker could exploit this shortcoming
with a \emph{masking attack}:
If an attacker can generate two
flows that are mapped to the same
LB,
the attacker could start by sending
a moderately bursty flow,
thereby occupy the LB, and initiate
the excessively bursty flow right 
afterwards. To mask
the excessively bursty flow,
the masking flow either has to be bursty
within the allowance or not send any
packets after it has been assigned
to the LB. 
Regarding accuracy,
the assignment
of a bursty flow to a free LB
is fairly unlikely, as the assignment
probability roughly
corresponds to the share of the pulsating flow
among all packets mapped to a given LB.
For a large number of flows or low overuse ratios, 
this probability may not suffice 
for swift detection.

To resolve this issue, 
\name{} employs a set~$\Pi$ of \emph{background counters} (BCs).
Each BC~$\pi \in \Pi$ is
assigned to an LB~$\lambda \in \Lambda$
and is tasked with finding the dominant
flow among the flows that are mapped to,
but not currently monitored by LB~$\lambda$.
This dominant
flow is identified with a probabilistic-decay
technique~\cite{yang2019heavykeeper,zhong2021burstsketch}.
When an LB is cleared,
the flow from the corresponding
BC is assigned to the LB directly,
and is evicted from the BC.
The background counters
mitigate the masking attack
with regular checks of the BC value. 
If this value exceeds a
\emph{push threshold}, 
the BC flow immediately replaces 
the corresponding LB flow.
Regarding accuracy, this background counting 
increases the probability that a bursty
flow is assigned to an LB, where it can then
be identified as excessively bursty.
In fact, background counting
significantly boosts the probability of assigning
the dominant flow to an LB,
and thus the recall of \name{} (cf. 
\ifthenelse{\boolean{fullpaper}}{\cref{sec:evaluation:prefiltering}}{Appendix B~\cite{scherrer2023albusfull}}).

Finally, \name{} requires an additional
safeguard to evict flows that finish
sending while being monitored 
by an LB. To eventually evict these flows, 
the timestamp in LBs is regularly checked
against a time-out.

\subsection{Description of \name{}}
\label{sec:design:algorithm}

In the following, we present a more detailed account
of the \name{} algorithm based on~\cref{fig:design:algorithm}. 
\name{} operates
with both an indexed set~$\Lambda$ of leaky buckets (LBs)
and an indexed set~$\Pi$ of background counters~(BCs),
where the associated LB~$\lambda$ and BC~$\pi$
are at the 
same index~$i$ of the respective sets, 
i.e., $\lambda = \Lambda[i]$
and~$\pi = \Pi[i]$. For any
packet with flow ID~$f$, size~$s$ and
timestamp~$t$, the index~$i$
is determined by a keyed hash function based
on the flow ID~$f$. After identifying
the pair~$(\lambda,\pi)$, 
\name{} performs the following case distinction.

If LB~$\lambda$ is still empty, flow~$f$
is assigned to~$\lambda$ by inserting
its flow ID~$f$ into~$\lambda$ and
setting the LB timestamp~$\lambda.t$ to the
packet timestamp~$t$ and the LB count~$\lambda.c$
to the packet size~$s$ (cf.~\cref{fig:design:algorithm} \circled{0}).
If LB~$\lambda$ already monitors flow~$f$,
then the leaky-bucket calculation is performed
to obtain the drain volume~$d$ and the new
LB count~$\lambda.c$. If~$\lambda.c$ exceeds
the burstiness allowance~$\beta$, 
flow~$f$ is reported and replaced by
the flow occupying BC~$\pi$ (cf.~\cref{fig:design:algorithm} \circled{1}). 
If the burstiness allowance is not
exceeded, flow~$\lambda.f$ is left
in LB~$\lambda$ (with updated
timestamp $\lambda.t$) if the packet size~$s$ 
is larger than the drain volume~$d$ 
(~\cref{fig:design:algorithm} \circled{2}) or replaced by
the flow from~$\pi$ otherwise 
(\circled{3}). When pulling
flow~$\pi.f$ into LB~$\lambda$,
the LB timestamp~$\lambda.t$ must be
set to~$-\infty$, as the timing of the
last packet of~$\pi.f$ is unknown
and the flow must not be overestimated.
BC~$\pi$ is cleared after
replacement.

So far, we have only considered cases where
an update of LB~$\lambda$ is directly possible.
However, if another flow occupies LB~$\lambda$,
BC~$\pi$ must be accessed. If BC~$\pi$
is empty, flow~$f$ is assigned to~$\pi$
by setting~$\pi.f$ to $f$ and $\pi.c$ to packet
size $s$ (cf. \cref{fig:design:algorithm} \circled{4}). 
If flow~$f$ already occupies BC~$\pi$, the
BC count~$\pi.c$ is increased by packet size~$s$
(cf. \cref{fig:design:algorithm} \circled{5}).
If~$\pi$ is already
occupied by another flow, 
\name{} employs a probabilistic-decay
technique: With a configurable probability $0.1^r$,
the BC count~$\pi.c$ is decremented by
the current packet size~$s$.
If~$\pi.c$ is reduced below 0,
the current flow~$f$ replaces 
flow~$\pi.f$ in BC~$\pi$ (\cref{fig:design:algorithm} 
\circled{6}).
A high~$r$ slows BC
decrements and BC flow replacements;
hence, the parameter~$r$
corresponds to the \emph{rigidity} of the
background counters.

Finally,
the algorithm contains two safeguards,
namely safeguards against masking attacks
and inactive flows.
To counter masking attacks.
the BC  count~$\pi.c$ is checked 
when increasing~$\pi.c$. If this
BC count exceeds a configurable 
threshold~$T$, the current flow~$\pi.f$ is 
swapped with the flow from
the corresponding LB (cf. \cref{fig:design:algorithm}
\circled{7}).
While LB counts and BC counts are not
strictly comparable, \name{} adopts the LB count
as the BC count in order 
not to lose information (A swap in reverse direction
might cause false positives).
The safeguard against inactive flows
corresponds to an occasional check of the
LB timestamp~$\lambda.t$, which is
performed even if the flow ID~$f$
of the current packet is not currently
assigned to~$\lambda$. 
The safeguard procedure checks whether
the LB timestamp predates the current
time by more than a time-out, and if yes,
replaces the flow~$\lambda.f$ with
the flow~$\pi.f$ from the BC.
This time-out is chosen 
such that the LB has certainly
been fully drained since the last
packet mapped to the LB was sent,
i.e., the time-out duration 
is~$\beta/\gamma$. 


\section{Analysis}
\label{sec:analysis}


\subsection{Security Analysis}
\label{sec:analysis:security}

\name{} detects excessively bursty flows
which are assigned to an LB for a sufficiently
long duration before the bursts ends.
To evade detection. the sender of an excessively
bursty flow must thus avoid or delay the
LB assignment.

\textbf{Single-flow masking.}
To completely avoid LB assignment,
an adversary could launch a \emph{masking attack},
i.e., occupy an LB
with a flow and then send a large
traffic burst in another flow that
is mapped to the same LB, speculating that
the more damaging flow will not be
inserted into the already occupied LB.
Masking flows can be flows that either are
bursty within acceptable limits (i.e., $\lambda.c < \beta$,
but~$s > d$ consistently) or do not
send any packets while they occupy
an LB (so the eviction condition is never evaluated). 
\name{} counters such attacks in a 
two-fold manner. 
First, \name{} provides
the push-based transfer,
i.e., overrides the LB flow
with a flow that exceeds threshold~$T$
in the BC; hence, an excessively bursty 
flow might be monitored
even if the masking flow is also bursty.
Second, the time-out check 
prevents
silent masking flows,
as these flows would be quickly evicted.

\textbf{Multi-flow masking.}
These safeguards might be circumvented
if an attacker creates a high
number of flows that map to the same
BC as a bursty flow. Then, the probabilistic-decay 
mechanism might keep the BC count of the bursty
flow low (i.e., no LB insertion by threshold violation)
or even keep the flow out of the BC
(i.e., no LB insertion on time-out).
However,
our experimental investigation  
in~\cref{sec:evaluation}
confirms that an appropriately configured BC
maintains high recall
even under a high number of simultaneous bursts.

\textbf{Outside-LB flooding.}
Since the adversary is thus unable to keep
the excessively bursty flows out of an LB
indefinitely, most damage can be caused by
sending excess traffic while the bursty
flow is not yet assigned to an LB, i.e.,
if the flow is assigned to a BC or no counter
at all. However, this attacker strategy is
limited by the lower bounds on the probability
of BC assignment and push transfers 
(\ifthenelse{\boolean{fullpaper}}{\cref{sec:analysis:accuracy}}{Appendix A~\cite{scherrer2023albusfull}}).
Moreover, an attacker cannot
observe when the malicious flow enters or exits
a BC, and does not know whether the
flow is already monitored by an LB.
Hence, the attacker bursts must be sized
conservatively to avoid detection.
Even if the adversary knew
the exact time at which the flow is transferred
from the BC to the LB,
this time is brought forward as the attacker
burst gets larger and background filtering is sped up. 

\textbf{Reset exploitation.}
A further weakness of previous algorithms
is their reliance on regular resets.
These resets allow an attacker to send a burst
around the reset time. With this timing,
the burst volume is split across two intervals
such that the threshold is not exceeded
in either interval. Since \name{} does not rely
on fixed time windows or resets, it is not
susceptible to this evasion strategy.

\textbf{Small-burst attack.}
If the adversary creates an enormous 
number of flows, each containing small bursts
which are also expected from benign flows, 
it is fundamentally impossible
for a volume-based algorithm
to identify these attacker flows
without also reporting benign flows.
This fundamental impossibility applies
to previous monitoring algorithms
(Count- and CountMin-Sketch) as
well as to \name{}.

To counter such small-burst attacks,
functionality provided by the DDoS defense
system is required. Depending on the
attack, the attacker flows might share 
some packet attributes which allow to
combine the flows into a meaningful,
large-volume aggregate.
In that case, the DDoS defense system
might apply the monitoring algorithm
to measure aggregates instead of fivetuple-based
flows. 
In the longer term, small-burst
attacks might be effectively countered
by DDoS defense systems that restrict
flow creation. For example,
COLIBRI~\cite{giuliari2021colibri}
relies on source authentication
and paid per-flow reservations, 
which limits the number
of flows over which attack traffic
can be distributed. Given this restriction, the
adversary must create medium-sized bursts
in each flow, which are again detectable.

\subsection{Complexity Analysis}
\label{sec:analysis:complexity}

The following complexity analysis
discusses memory consumption and
processing overhead of \name{}.

\textbf{Memory consumption.}
\label{sec:analysis:complexity:memory}
The memory consumption of \name{} is linear in the
number~$|\Lambda| = |\Pi|$ of LB-BC pairs,
where each LB contains three fields
(flow ID, timestamp, and count) and each BC 
contains two fields (flow ID and count).
For a realistic traffic profile, each LB-BC pair 
can be represented by 16 bytes, which is justified 
in \ifthenelse{\boolean{fullpaper}}{\cref{sec:memory-analysis}}{Appendix C~\cite{scherrer2023albusfull}}.

This low memory consumption ensures
high memory efficiency: \cref{sec:evaluation:sensitivity}
suggests that a memory consumption of 300KB
(on the order of usual L1/L2 cache sizes)
allows \name{} to outperform previous algorithms
(when considering both recall and precision)
on a 10~Gbps link~\cite{caida201810}.

\textbf{Processing overhead.}
\label{sec:analysis:complexity:processing}
Unlike its competitor schemes, 
the processing overhead per packet
in \name{} is independent of detector
memory. This independence stems
from \name{}'s avoidance of list
iterations, which are
used in some monitoring algorithms
such as BurstSketch (when populating or
analyzing the Snapshot component~\cite{zhong2021burstsketch})
and EARDet (when looking for an empty
counter for flow insertion~\cite{Wu14}).
Moreover, unlike BurstSketch and EARDet,
\name{} does not require associative arrays 
(i.e., hashmaps), which are challenging to 
implement in  a lightweight manner in hardware.
Moreover, \name{} requires exactly 1 hash operation,
whereas other schemes like Count-Min Sketch~\cite{Cormode2005},
CountSketch~\cite{charikar2002finding},
or BurstSketch~\cite{zhong2021burstsketch} 
use multiple hash operations. 
Furthermore,  the most complex arithmetic
computation, namely the 
LB update, is
only performed in the rare case where the flow
is monitored by the LB. 
The lightweight update procedure
thus allows high-speed processing, which we confirm
with an FPGA implementation using 5 ns per packet 
(cf.~\cref{sec:implementation}).
\section{Implementation}
\label{sec:implementation}

First, we implemented \name{} for an FPGA to demonstrate
its suitability for high-speed packet processing (cf.~\cref{sec:implementation:fpga}). 
Second, we implemented~\name{} in P4, 
showing its suitability for programmable 
switches (cf.~\cref{sec:implementation:p4}).

\subsection{FPGA Implementation} 
\label{sec:implementation:fpga}

To demonstrate that \name{} allows
efficient packet processing,
we implemented \name{} for a
Xilinx FPGA.

\begin{figure}
    \centering
    \begin{tikzpicture}[
flowsquare/.style={draw=black!60, shading=radial,outer color={rgb,255:red,137;green,207;blue,240},inner color=white, thick, minimum size=\nodewidth,outer sep=0pt},
flowsizesquare/.style={draw=black!60, inner color=white, thick, minimum size=\nodewidth},
indexsquare/.style={draw=black!60,  shading=radial,outer color={rgb,1:red,0.8;green,0.8;blue,0.8},inner color=white, thick,minimum width=\counterwidth,minimum height=\indexheight,align=center},
lbsquare/.style={draw=black!60, shading=radial,outer color={rgb,255:red,240;green,207;blue,137},inner color=white, thick,minimum width=\counterwidth,minimum height=\lbheight,align=center},
pcsquare/.style={draw=black!60, shading=radial,outer color={rgb,1:red,0.53;green,0.66;blue,0.42},inner color=white, thick,minimum width=\counterwidth,minimum height=\lbheight,align=center},
]

    \def\nodewidth{6mm};
    \def\counterwidth{6mm};
    \def\indexheight{5mm};
    \def\lbheight{6mm};
    
    \node[flowsquare] (f1) at (0,0.5) {$f_1$};
    
    \node[indexsquare,align=center] (xoodoo) at (1.5, 0.5) {Xoodoo-\\NC hash};
    \draw[-latex] (f1.east) -- (xoodoo.west) 
    node[midway,yshift=10mm,rotate=90] {Xoodoo-NC};
    \draw [decorate,decoration={brace,amplitude=5pt}]
(xoodoo.north west) -- (xoodoo.north east) node [black,midway,yshift=10pt] {96 bits};
    \node[draw=red,dashed,thick,minimum width=5mm,minimum height=10mm] (xoodoo_lsb) at ($(xoodoo) + (0.5, 0)$) {};
    \draw [decorate,decoration={brace,amplitude=5pt,mirror}]
(xoodoo_lsb.south west) -- (xoodoo_lsb.south east) node [black,midway,yshift=-10pt] {10 bits};

    \node[indexsquare,align=center] (onehot) at (3.5, 0.5) {One-hot\\vector};
    \draw[-latex] (xoodoo_lsb.east) -- (onehot.west)
    node[black,midway,yshift=7mm,rotate=90] {Decode};
    \draw [decorate,decoration={brace,amplitude=5pt}]
(onehot.north west) -- (onehot.north east) node [black,midway,yshift=10pt] {1024 bits};

    \node[lbsquare] at (5.5,1.5) (l0) {$\lambda_0$};
    \node[lbsquare] at (5.5,0.7) (l1) {$\lambda_1$};
    \node[lbsquare] at (5.5,-0.2) (lC) {$\lambda_{C}$};
    
    \node[pcsquare] at (6.15,1.5) (p0) {$\pi_0$};
    \node[pcsquare] at (6.15,0.7) (p1) {$\pi_1$};
    \node[pcsquare] at (6.15,-0.2) (pC) {$\pi_{C}$};
    
    \node[] (cell_center) at (5.825, 0.25) {...};
    
    \draw[-latex] (onehot.east) -- (l1.west);
    
    \definecolor{antiquefuchsia}{rgb}{0.57, 0.36, 0.51}
    \node[draw=antiquefuchsia,dotted,ultra thick,minimum width=15mm,minimum height=8mm] (cell_square) at ($(cell_center) + (0.0,1.25)$) {};
    \node (cell_desc) at ($(cell_square) + (-13mm,+3mm)$) {\textcolor{antiquefuchsia}{Cell 0}};
    
     \draw [decorate,decoration={brace,amplitude=8pt,mirror,aspect=0.85}]
(l0.north west) -- (lC.south west) node [black,midway,xshift=-6mm,yshift=-8mm,align=center] {1024\\cells};
    
    \node (and_gate_entry) at ($(cell_center) + (1.3,0.25)$) {};
    \draw[-] ($(and_gate_entry)+(0,0.5)$) -- ($(and_gate_entry)-(0,0.5)$);
    \draw[-] ($(and_gate_entry)+(0,0.5)$) .. controls ($(and_gate_entry)+(1.5,0.5)$) and ($(and_gate_entry)+(1.5,-0.5)$)  .. ($(and_gate_entry)-(0,0.5)$);
    \node (and_label) at ($(and_gate_entry)+(0.5,0.0)$) {AND};
    
    \draw (p0.east) -- (and_gate_entry.center);
    \draw (p1.east) -- (and_gate_entry.center);
    \draw (pC.east) -- (and_gate_entry.center);
\end{tikzpicture}
    \vspace{-15pt}
    \caption{FPGA implementation design ($C = 1024 - 1$).}
    \label{fig:implementation:design}
\end{figure}
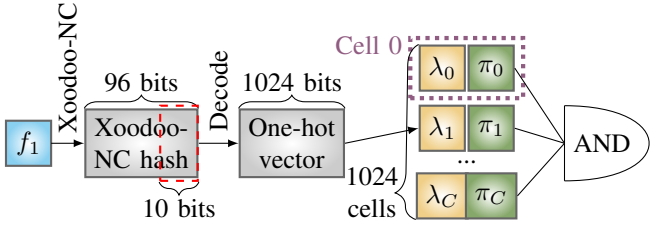

\textbf{Design.}
\label{sec:implementation:design}
\Cref{fig:implementation:design} illustrates
the design of our \name{} FPGA implementation. 
The implementation
starts by hashing the flow ID of an incoming
packet (e.g., flow
fivetuple) with the Xoodoo-NC hash 
function~\cite{sateesan2020novel},
which is a non-cryptographic hash function 
based on consecutive Xoodoo permutations~\cite{daemen2018xoodoo}.
From the 96-bit Xoodoo-NC digest, 
the implementation decodes the 10 least
significant bits into a one-hot
vector of 1024 bits. 
In contrast to a decoding design, 
a single memory was used. 
Every binary-encoded address stores 
a corresponding one-hot encoded value.
The single positive bit
in this vector then triggers
the activation of the corresponding 
\emph{cell}, i.e., an LB-BC pair (cf.~\Cref{fig:implementation:cell}). 
\begin{figure}
    \centering
    \includegraphics[width=0.9\linewidth]{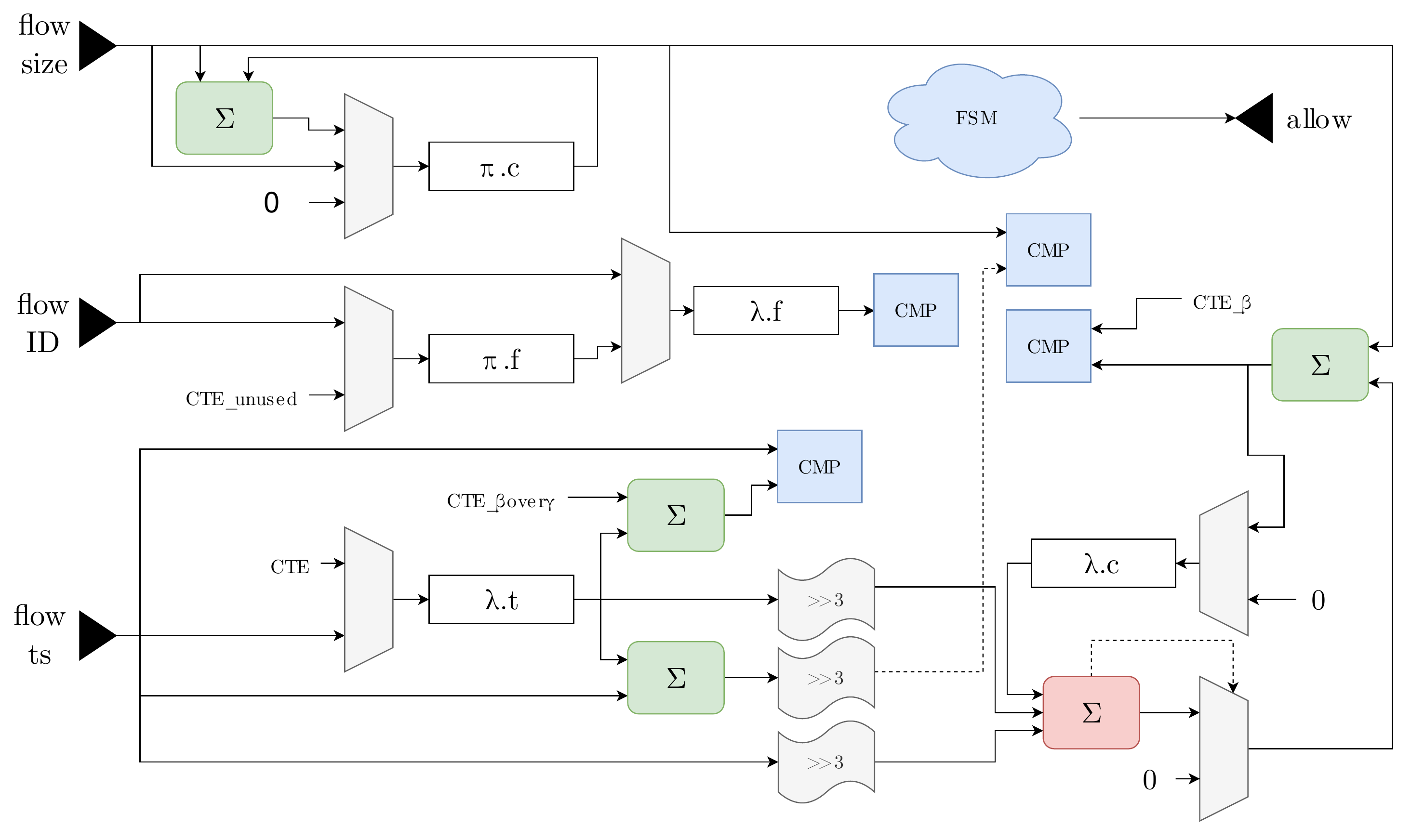}
    \vspace{-10pt}
    \caption{A single cell, consisting of ripple-carry adders (green), a carry-save adder (red), and the decision-making logic (blue).}
    \label{fig:implementation:cell}
\end{figure}


\textbf{Resource evaluation.}
\label{sec:implementation:evaluation}
We evaluate our implementation
on a Xilinx Virtex UltraScale+ 
FPGA VCU118~\cite{xilinx2021} with
a dual 112Gbps network interface.
On this board, our implementation 
achieves a clock period of 5.0 nanoseconds,
which corresponds to a packet-processing
rate of 200 million packets per second
for packets of 64 bytes.
With these minimum-sized IP packets,
this packet rate corresponds to
a processing capacity of 102Gbps,
which is compatible with the network interface
of the FPGA. For a realistic
traffic mix, such as the IMIX traffic
distribution with an average packet
size of 350 bytes~\cite{popoviciu2007}, 
the packet rate would allow to process
a traffic stream of 560 Gbps, but
the network interface then constitutes
the bottleneck.
Given such a traffic profile, the forwarding
switch could asynchronously provide
the \name{} FPGA with
pure packet metadata, which amounts
to maximum 39 bytes per packet
or 62.4 Gbps of inbound traffic
at the FPGA network interface.

The resource utilization of the
implementation on the Xilinx FPGA
amounts to 40\% of available
look-up tables (LUT),
20\% of flip-flops (FF),
and 2\% of BRAM.

\subsection{P4 Implementation}
\label{sec:implementation:p4}

We created an additional implementation in
P4~\cite{bosshart2014p4}, a programming language
for programmable 
switches~\cite{agrawal2020intel}.
Our P4 code is available online~\cite{albusp4code} 
and can be tested using the virtual V1Model 
switch~\cite{v1model}. 
\ifthenelse{\boolean{fullpaper}}{\Cref{sec:implementation:p4-details}}{Appendix D~\cite{scherrer2023albusfull}} 
presents a conceptual overview of the implementation design.
\section{Evaluation}
\label{sec:evaluation}

To compare \name{}
to its competitor algorithms CountMin-Sketch,
CountSketch, and BurstSketch
under a variety of parameters, we perform
a sensitivity analysis of all
algorithms in~\cref{sec:evaluation:sensitivity}.
We also investigate the effectiveness
of background filtering in terms of 
accuracy 
(\ifthenelse{\boolean{fullpaper}}{\cref{sec:evaluation:prefiltering}}{Appendix B~\cite{scherrer2023albusfull}}). 

\subsection{Evaluation Set-Up}
\label{sec:evaluation:setup}

For our evaluation, we use a simulation framework
implemented in Go. This framework replays a
CAIDA trace from a 10Gbps link~\cite{caida201810} as background traffic.
This trace is augmented with a simulated burst-flood attack
such that the total rate exceeds the link capacity.
These burst-flood attacks vary in the overuse ratio of bursts,
the burst duration, and the number of bursts during the 
observation interval (5 seconds).
We measure the detection accuracy of the
monitoring algorithms by means of \emph{recall}
and \emph{precision}~(cf.~\cref{sec:problem:definition}).

Every experiment in this section is run 6 times. 
However, since most results are highly stable and standard deviations are small,
the error bars are barely visible.

\subsection{Comparative Sensitivity Analysis}
\label{sec:evaluation:sensitivity}

In this section, we investigate the detection accuracy
of \name{} in different scenarios, and simultaneously
compare it to the monitoring algorithms used in DDoS defense systems,
i.e., CountMin-Sketch~\cite{Cormode2005} (used by Poseidon~\cite{zhang2020poseidon}
and Ripple~\cite{xing2021ripple}),
and the CountSketch~\cite{charikar2002finding} (used by Jaqen~\cite{liu2021jaqen}).
Since these algorithms are subject to a recall-precision trade-off due
to their landmark-window model, they are evaluated for multiple threshold factors (cf.~\cref{sec:problem:motivation}).
To distinguish versions of the algorithm with different threshold factors,
we write \textit{a(k)} for algorithm~\textit{a} with threshold factor~\textit{k}.
In addition, we also evaluate BurstSketch~\cite{zhong2021burstsketch},
which is a specialized monitoring algorithm for burst detection.

In the comparative analysis, all detectors are
equipped with the same amount of memory.
The competitor schemes
have been optimally according to
the respective papers. \name{} has been configured
with an optimal rigidity of~$r=0$ and~$T=10\mathrm{KB}$ 
(cf. \ifthenelse{\boolean{fullpaper}}{\cref{sec:evaluation:prefiltering}}{Appendix B~\cite{scherrer2023albusfull}}).

\newcommand{\evalplotwidth}{0.187}
\begin{figure*}[t]
    \centering
    \begin{minipage}{\evalplotwidth\linewidth}
    \captionsetup{justification=centering}
    \begin{subfigure}{\linewidth}
        \centering
        \includegraphics[width=\linewidth,trim=0 5 0 0]{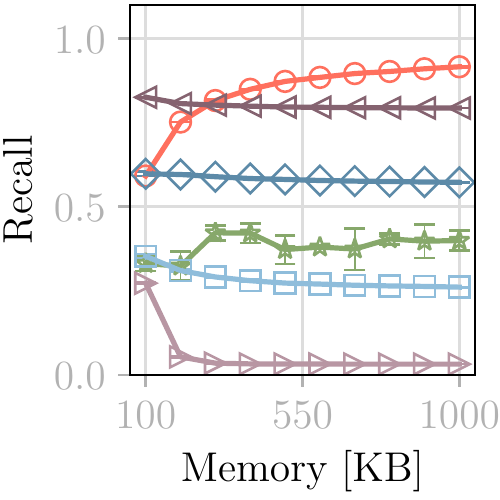}
        \caption{Recall.}
        \label{fig:evaluation:sensitivity:memory:recall}
    \end{subfigure}\\
    \begin{subfigure}{\linewidth}
        \centering
        \includegraphics[width=\linewidth,trim=0 5 0 0]{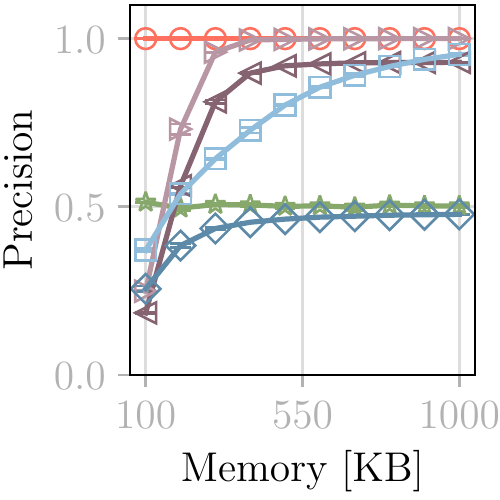}
        \caption{Precision.}
        \label{fig:evaluation:sensitivity:memory:precision}
    \end{subfigure}
    \caption{\\Memory.}
    \label{fig:evaluation:sensitivity:memory}
    \end{minipage}\ \vrule\ 
    \begin{minipage}{\evalplotwidth\linewidth}
    \captionsetup{justification=centering}
    \begin{subfigure}{\linewidth}
        \centering
        \includegraphics[width=\linewidth,trim=0 5 0 0]{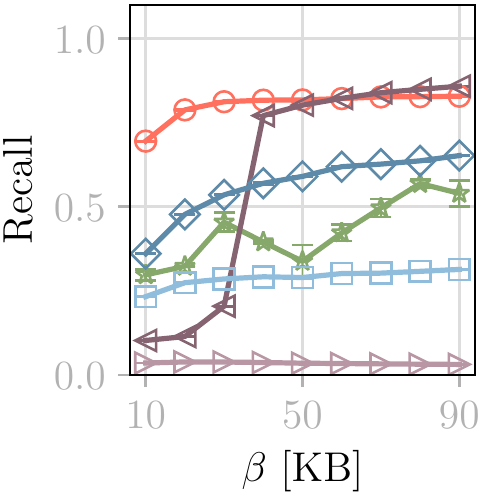}
        \caption{Recall.}
        \label{fig:evaluation:sensitivity:beta:recall}
    \end{subfigure}\\
    \begin{subfigure}{\linewidth}
        \centering
        \includegraphics[width=\linewidth,trim=0 5 0 0]{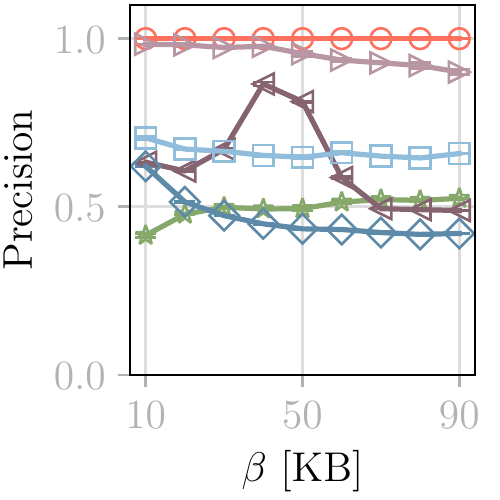}
        \caption{Precision.}
        \label{fig:evaluation:sensitivity:beta:precision}
    \end{subfigure}
    \caption{\\Allowance~$\beta$.}
    \label{fig:evaluation:sensitivity:beta}
    \end{minipage}
    \vrule
    \ \begin{minipage}{\evalplotwidth\linewidth}
    \captionsetup{justification=centering}
    \begin{subfigure}{\linewidth}
        \centering
        \includegraphics[width=\linewidth,trim=0 5 0 0]{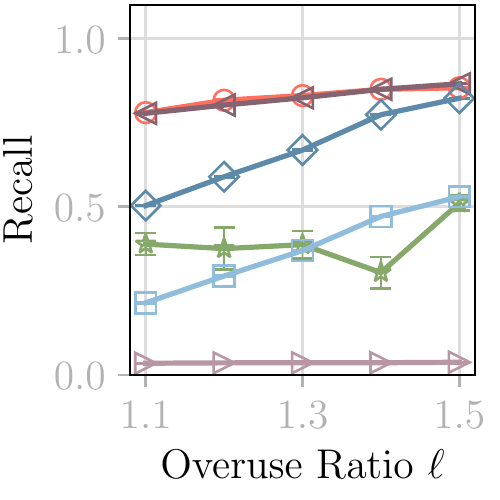}
        \caption{Recall.}
        \label{fig:evaluation:sensitivity:or:recall}
    \end{subfigure}\\
    \begin{subfigure}{\linewidth}
        \centering
        \includegraphics[width=\linewidth,trim=0 5 0 0]{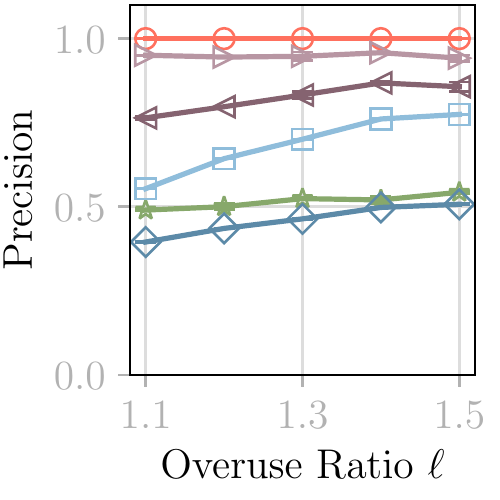}
        \caption{Precision.}
        \label{fig:evaluation:sensitivity:or:precision}
    \end{subfigure}
    \caption{\\Overuse ratio.}
    \label{fig:evaluation:sensitivity:or}
    \end{minipage}
    \vrule\ 
    \begin{minipage}{\evalplotwidth\linewidth}
    \captionsetup{justification=centering}
    \begin{subfigure}{\linewidth}
        \centering
        \includegraphics[width=\linewidth,trim=0 5 0 0]{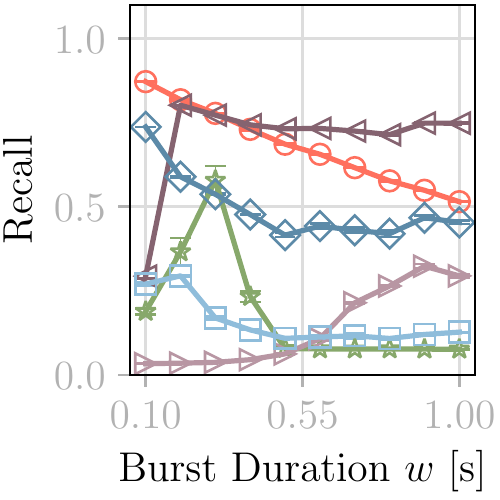}
        \caption{Recall.}
        \label{fig:evaluation:sensitivity:pd:recall}
    \end{subfigure}\\
    \begin{subfigure}{\linewidth}
        \centering
        \includegraphics[width=\linewidth,trim=0 5 0 0]{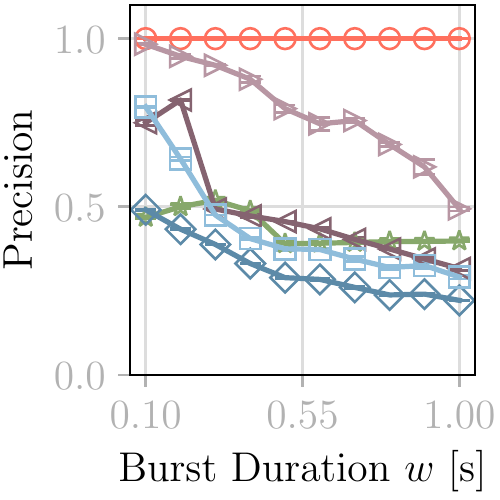}
        \caption{Precision.}
        \label{fig:evaluation:sensitivity:pd:precision}
    \end{subfigure}
    \caption{\\Burst duration.}
    \label{fig:evaluation:sensitivity:pd}
    \end{minipage}
    \vrule\ 
    \begin{minipage}{\evalplotwidth\linewidth}
    \captionsetup{justification=centering}
    \begin{subfigure}{\linewidth}
        \centering
        \includegraphics[width=\linewidth,trim=0 5 0 0]{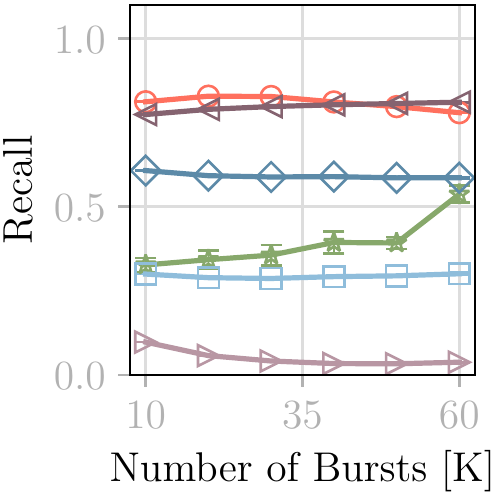}
        \caption{Recall.}
        \label{fig:evaluation:sensitivity:of:recall}
    \end{subfigure}\\
    \begin{subfigure}{\linewidth}
        \centering
        \includegraphics[width=\linewidth,trim=0 5 0 0]{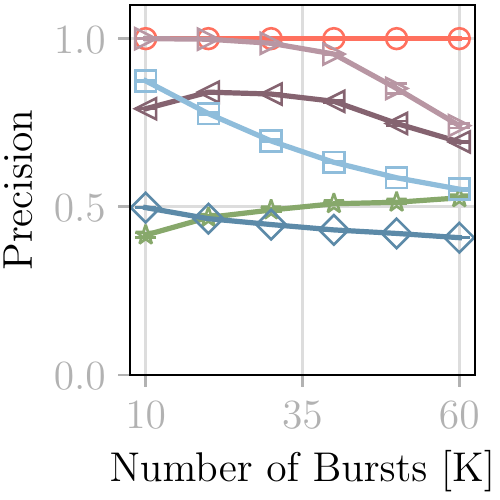}
        \caption{Precision.}
        \label{fig:evaluation:sensitivity:of:precision}
    \end{subfigure}
    \caption{\\Number of bursts.}
    \label{fig:evaluation:sensitivity:of}
    \end{minipage}\\\ContinuedFloat
    \begin{subfigure}{\linewidth}
        \centering
        \includegraphics[width=\linewidth,trim=1 50 1 620,clip]{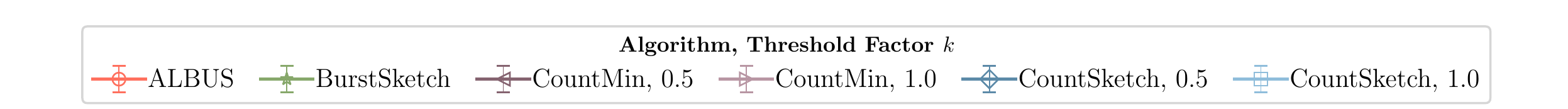}
    \end{subfigure}
\end{figure*}

\textbf{Base configuration.} The starting point of our sensitivity analysis is
a base configuration of experiment parameters.
In the base configuration,
every algorithm obtains a memory allocation of 300KB,
which is on the order of modern L1/L2 cache sizes.
All algorithms try to enforce a flow specification with
rate~$\gamma = 1$~Mbps and burstiness allowance~$\beta = 50$~KB,
allowing a flow to send at $1+0.4/w$~Mbps over~$w$ seconds.
Only~$\sim1$\% of background flows from the 
CAIDA trace violate this specification.
The attack in the base configuration contains bursts with 
an overuse ratio of~$\ell = 1.2$ and a burst width of 200~ms, 
i.e., bursts of rate~$3.4$~Mbps during 200~ms. 
To create an aggregate attack rate of around 5~Gbps over 5~seconds,
38,000 bursts are distributed across the observation interval uniformly at random.
In the following, we discuss how
changes in a single parameter affect the detection accuracy
of all monitoring algorithms.

\textbf{Memory.} \cref{fig:evaluation:sensitivity:memory}
illustrates how recall and precision are affected
as the memory allocation of the monitoring algorithms grows
from 100~KB to 1~MB. While CountSketch and
CountMin-Sketch can utilize the additional memory to increase
precision, their recall surprisingly \emph{decreases} with additional
memory. We observe this effect because additional memory is added
in the form of additional counters, and thus the number of flows
sharing a counter decreases with increasing memory.
A low number of flows per counter makes threshold violation
less likely, decreasing recall and increasing precision.
For BurstSketch, both recall and precision are consistently low,
pointing to a mismatch between the detection capabilities
of BurstSketch and the objective in the experiment.
Finally, \name{} benefits from additional memory in terms
of recall, whereas it achieves perfect precision by design.
From a memory allocation of 300KB, \name{} is superior
to all other algorithms regarding both recall and precision.
Other algorithms can beat \name{} 
in one metric only at the cost of bad performance in the other metric, e.g.,
at 100~KB, CountMin-Sketch(0.5) outperforms \name{}
regarding recall by 0.2, but only with a precision that is 5 times lower.

\textbf{Flow specification.}
In \cref{fig:evaluation:sensitivity:beta}, the burstiness
allowance~$\beta$ of the flow specification is varied.
Hence, this experiment shows how monitoring algorithms behave
when the allowed burst volume~$\beta$ grows compared to the 
allowed base rate~$\gamma$. Recall increases with~$\beta$ for all monitoring algorithms,
mostly because the bursts in the simulated attack
are dimensioned relative to~$\beta$, and thus become larger and
easier to detect. Precision, however, generally decreases with~$\beta$
for CountSketch and CountMin-Sketch, because the number of false positives
stays roughly constant whereas the number of true positives
decreases (i.e., fewer specification-violating flows in the background traffic).
In general, \name{} is uniquely effective at enforcing
tight flow specifications, i.e., a low burstiness allowance.
Moreover, \name{} benefits from loosening the specification,
whereas the performance of competitor schemes under varying flow
specifications is hard to predict.

\textbf{Overuse ratio.}
\cref{fig:evaluation:sensitivity:or} visualizes the detection performance
of the algorithms under varying overuse ratios.
Unsurprisingly, higher overuse ratios generally lead to higher recall
because larger bursts are easier to detect. Since the number of correctly detected
bursts grows compared to the number of incorrectly reported bursts,
precision also rises together with the overuse ratio.
\name{} is on par with CountMin-Sketch(0.5)
regarding recall, but is consistently more precise. 

\textbf{Burst width.} 
The variation of burst width in \cref{fig:evaluation:sensitivity:pd}
underlines that sketches are inflexible with respect to burst
width, i.e., they have a distinct peak in recall around a certain
burst width. In contrast, the decreasing recall of~\name{} for growing
burst width is due to the constant overuse ratio: As the burst width
grows, a constant excess burst volume is distributed over a longer
time, lowering the burst rate to which~\name{} is sensitive.
Nonetheless, the precision of~\name{} is stable for varying burst width,
whereas precision starkly decreases for the sketch algorithms.
The reason for this decrease is subtle: As bursts get longer,
they more often share counters with other flows,
increasing false positives.

\begin{figure}[t]
    \centering
    \begin{minipage}{0.85\linewidth}
        \captionsetup{justification=centering}
    \begin{subfigure}{0.49\linewidth}
        \centering
        \includegraphics[width=\linewidth,trim=0 5 0 0]{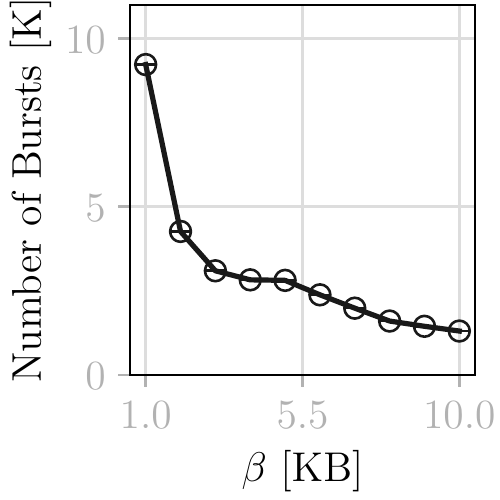}
        \caption{Excessive bursts.}
        \label{fig:evaluation:cicd:positives}
    \end{subfigure}
    \begin{subfigure}{0.49\linewidth}
        \centering
        \includegraphics[width=\linewidth,trim=0 5 0 0]{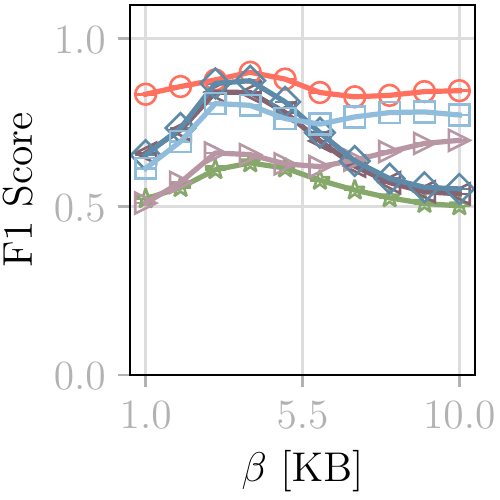}
        \caption{F1 score.}
        \label{fig:evaluation:cicd:f1}
    \end{subfigure}
    \end{minipage}
    \caption{CIC-DDoS2019 trace.}
    \label{fig:evaluation:cicd}
\end{figure}

\textbf{Number of bursts.}
In \cref{fig:evaluation:sensitivity:of}, the number of bursts in
the observation period is grown from 10,000 to 60,000, corresponding
to a variation in aggregate attack rate from 1.36~Gbps to 8.16~Gbps.
The recall for all detectors is mostly stable across the investigated range.
Only BurstSketch achieves a higher recall for higher burst numbers;
an inspection of the algorithm suggests that a higher number of flows
increases the contention in the first-stage component of BurstSketch,
which intensifies filtering such that the second-stage component
is less congested and contains more truly large flows. 
Regarding precision, we again observe that the CountSketch and the CountMin-Sketch
perform worse under high load, again because more intense counter sharing
leads to more false positives. CountMin-Sketch(0.5), which is on par with
\name{} in terms of recall, is at least 20\% less precise.

\textbf{Attack type.} Finally, we consider a completely different type 
of attack traffic by using the CIC-DDoS2019 dataset~\cite{sharafaldin2019developing}.
Specifically, we replay the excerpt containing
a UDP flood, and instruct all algorithms to detect excessively
large bursts for a range of flow specifications. 
Since the aggregate rate in the CIC-DDoS dataset is
far lower than in the CAIDA dataset, we investigate flow specifications
for~$\gamma = 0.1$~Mbps and~$\beta \in [1\mathrm{~KB}, 10\mathrm{~KB}]$.
The CIC-DDoS attack traffic is notably different from the
burst-flood attacks considered before, as it contains only
a few thousand bursts over 24 minutes
(\cref{fig:evaluation:cicd:positives}).
Nonetheless, \name{} consistently outperforms remaining
algorithms in $F_1$ score (\cref{fig:evaluation:cicd:f1}),
demonstrating that \name{} is effective beyond 
burst-flood attacks.

\textbf{Overall performance.} In summary, \name{}
outperforms its competitor algorithms over a wide range of
scenarios, considering detector configurations, attack properties,
and background traffic 
(cf. \ifthenelse{\boolean{fullpaper}}{\cref{sec:syn-evaluation}}{Appendix E~\cite{scherrer2023albusfull}}). 
The improvement by~\name{} is especially strong for short bursts
and tight flow specifications (i.e., low~$\beta$ compared to~$\gamma$).
While the sketch algorithms can be configured such that they are
competitive with \name{} in one metric, such a configuration
leads to poor performance in the other metric.
For example, CountMin-Sketch(0.5) is frequently competitive with~\name{}
in terms of recall, but is consistently less precise.
Conversely, CountMin-Sketch(1.0) achieves competitive precision,
but inferior recall.

\section{Related Work}

In this section, we discuss previous probabilistic monitoring algorithms
and evaluate their suitability for burst detection. 

Since keeping track of each individual flow is impractical,
numerous probabilistic flow-monitoring
algorithms have been proposed, most
prominently  
CountMin-Sketch~\cite{Cormode2005} 
and Count-Sketch~\cite{charikar2002finding}.
However, as we have shown in~\cref{sec:problem:motivation},
these algorithms suffer from poor accuracy in the evaluated attacks
because they rely on discrete time windows.
While the inaccuracy can be partly remedied
by retrospective sketch-analysis techniques
such as SeqSketch~\cite{huang2021toward}, PR-Sketch~\cite{sheng2021pr},
or LOFT~\cite{scherrer2021lowrate},
the issues of discrete time windows and false positives
remain. Similar issues plague other
sketches~\cite{Liu2016,yang2018elastic,zhong2021burstsketch}, 
and top-$k$ detection schemes~\cite{sivaraman2017heavy,yang2019heavykeeper}.
EARDet~\cite{Wu14} does not rely on resets
and yields no false positives,
but performs expensive list iterations for every packet. 
The monitoring algorithm
in ACC-Turbo~\cite{alcoz2022aggregate} is based on the
assumption that attack flows can be correlated based
on header information; we make no such assumption.

\section{Conclusion}


In this work, 
we demonstrate that the sketch algorithms used in 
modern DDoS defense system provide poor
detection accuracy under burst-flood attacks, as they
fail to report the set of excessively bursty
flows without false inclusions.
For example, given a flood of
bursts that last 500~ms and exceed the 
allowed burst volume by 20\%,
the CountMin-Sketch~\cite{Cormode2005}
detects either 75\% of allowance-violating
flows with a false-positive rate of 50\%,
or 1\% of allowance-violating flows
with a false-positive rate of 1\%,
depending on the configuration.
The source of this problem is that
sketch algorithms need to regularly
reset their data structure, which
conflicts with arbitrary timing and duration
of bursts.

Our algorithm, \name{}, considerably improves
upon previous algorithms by continuously selecting
the subset of flows that are precisely monitored 
within limited memory.
Thanks to its reliance on the leaky-bucket
algorithm, its avoidance of hard resets,
and its leverage of filtering techniques,
\name{} does not falsely report any allowance-conforming flows, 
but maintains high recall.
In our experiments, \name{} frequently 
outperforms the other investigated algorithms in
\emph{both} precision and recall, and
consistently outperforms them in one of these metrics. 
Regarding processing efficiency,
\name{} consistently avoids design primitives
that prevent efficient hardware
implementation,
and thus enables an FPGA implementation
which can process
200 million packets per second.
Thanks to its frugality, \name{} 
can also be readily implemented in P4, in turn enabling simple integration
with 
programmable switches.

In summary, \name{} represents an 
important complement to the monitoring algorithms
in current DDoS defense systems,
as \name{} compensates the weaknesses of these algorithms
under pessimal workloads. In future work,
we will identify optimal combinations
of these traditional monitoring algorithms 
with \name{} to achieve comprehensive DDoS mitigation.


\section*{Acknowlegements}
We gratefully acknowledge support from ETH Zurich, and from SNSF
(200021L\_182005) and FWO (G0E0719N) for project ESCALATE.
Moreover, we thank Piet De Vaere, Marc Wyss, Jonghoon Kwon,
and the anonymous reviewers for their helpful feedback.

\newpage
\bibliographystyle{IEEEtran}
\bibliography{ESCALATE}

\begin{thebibliography}{10}
\providecommand{\url}[1]{#1}
\csname url@samestyle\endcsname
\providecommand{\newblock}{\relax}
\providecommand{\bibinfo}[2]{#2}
\providecommand{\BIBentrySTDinterwordspacing}{\spaceskip=0pt\relax}
\providecommand{\BIBentryALTinterwordstretchfactor}{4}
\providecommand{\BIBentryALTinterwordspacing}{\spaceskip=\fontdimen2\font plus
\BIBentryALTinterwordstretchfactor\fontdimen3\font minus
  \fontdimen4\font\relax}
\providecommand{\BIBforeignlanguage}[2]{{%
\expandafter\ifx\csname l@#1\endcsname\relax
\typeout{** WARNING: IEEEtran.bst: No hyphenation pattern has been}%
\typeout{** loaded for the language `#1'. Using the pattern for}%
\typeout{** the default language instead.}%
\else
\language=\csname l@#1\endcsname
\fi
#2}}
\providecommand{\BIBdecl}{\relax}
\BIBdecl

\bibitem{zhang2020poseidon}
M.~Zhang, G.~Li, S.~Wang, C.~Liu, A.~Chen, H.~Hu, G.~Gu, Q.~Li, M.~Xu, and
  J.~Wu, ``Poseidon: Mitigating volumetric {DDoS} attacks with programmable
  switches,'' in \emph{the 27th Network and Distributed System Security
  Symposium (NDSS 2020)}, 2020.

\bibitem{xing2021ripple}
J.~Xing, W.~Wu, and A.~Chen, ``Ripple: A {Programmable, Decentralized
  Link-Flooding Defense Against Adaptive Adversaries},'' in \emph{30th USENIX
  Security Symposium (USENIX Security 21)}, 2021, pp. 3865--3881.

\bibitem{liu2021jaqen}
Z.~Liu, H.~Namkung, G.~Nikolaidis, J.~Lee, C.~Kim, X.~Jin, V.~Braverman, M.~Yu,
  and V.~Sekar, ``Jaqen: A {High-Performance} {Switch-Native} approach for
  detecting and mitigating volumetric {DDoS} attacks with programmable
  switches,'' in \emph{30th USENIX Security Symposium (USENIX Security 21)},
  2021, pp. 3829--3846.

\bibitem{giuliari2021colibri}
G.~Giuliari, D.~Roos, M.~Wyss, J.~A. Garcia-Pardo, M.~Legner, and A.~Perrig,
  ``Colibri: A {C}ooperative {L}ightweight {I}nter-domain
  {B}andwidth-{R}eservation {I}nfrastructure,'' \emph{Proceedings of {ACM}
  {C}o{NEXT}}, 2021.

\bibitem{alcoz2022aggregate}
A.~G. Alcoz, M.~Strohmeier, V.~Lenders, and L.~Vanbever, ``Aggregate-based
  congestion control for pulse-wave {DDoS} defense,'' in \emph{Proceedings of
  the ACM SIGCOMM 2022 Conference}, 2022, pp. 693--706.

\bibitem{Cormode2005}
\BIBentryALTinterwordspacing
G.~Cormode and S.~Muthukrishnan, ``{An Improved Data Stream Summary: The
  Count-Min Sketch and its Applications},'' \emph{Journal of Algorithms},
  vol.~55, no.~1, pp. 58--75, 2005. [Online]. Available:
  \url{http://linkinghub.elsevier.com/retrieve/pii/S0196677403001913}
\BIBentrySTDinterwordspacing

\bibitem{charikar2002finding}
M.~Charikar, K.~Chen, and M.~Farach-Colton, ``Finding frequent items in data
  streams,'' in \emph{International Colloquium on Automata, Languages, and
  Programming}, 2002.

\bibitem{sharafaldin2019developing}
I.~Sharafaldin, A.~H. Lashkari, S.~Hakak, and A.~A. Ghorbani, ``Developing
  realistic distributed denial of service (ddos) attack dataset and taxonomy,''
  in \emph{2019 International Carnahan Conference on Security Technology
  (ICCST)}.\hskip 1em plus 0.5em minus 0.4em\relax IEEE, 2019, pp. 1--8.

\bibitem{ddosguard2019pulsewave}
DDoS-GUARD, ``Hidden threat of {P}ulse {W}ave {DD}o{S} attacks,''
  \url{https://ddos-guard.net/en/info/blog-detail/hidden-threat-of-pulse-wave-ddos-attacks},
  2019.

\bibitem{zeifman2017attackers}
I.~Zeifman, ``Attackers {U}se {DD}o{S} {P}ulses to {P}in {D}own {M}ultiple
  {T}argets,''
  \url{https://www.imperva.com/blog/pulse-wave-ddos-pins-down-multiple-targets/},
  2017.

\bibitem{rasti2015temporal}
R.~Rasti, M.~Murthy, N.~Weaver, and V.~Paxson, ``Temporal lensing and its
  application in pulsing denial-of-service attacks,'' in \emph{2015 IEEE
  Symposium on Security and Privacy}.\hskip 1em plus 0.5em minus 0.4em\relax
  IEEE, 2015, pp. 187--198.

\bibitem{rossow2014amplification}
C.~Rossow, ``Amplification hell: Revisiting network protocols for {DDoS}
  abuse.'' in \emph{NDSS}, 2014.

\bibitem{griffioen2021scan}
H.~Griffioen, K.~Oosthoek, P.~van~der Knaap, and C.~Doerr, ``Scan, test,
  execute: Adversarial tactics in amplification {DDoS} attacks,'' in
  \emph{Proceedings of the 2021 ACM SIGSAC Conference on Computer and
  Communications Security}, 2021, pp. 940--954.

\bibitem{agrawal2020intel}
A.~Agrawal and C.~Kim, ``Intel {Tofino2-A 12.9 Tbps P4-Programmable Ethernet
  Switch}.'' in \emph{Hot Chips Symposium}, 2020, pp. 1--32.

\bibitem{Netflow}
\BIBentryALTinterwordspacing
B.~Claise, ``{Cisco Systems NetFlow Services Export Version 9},'' RFC 3954
  (Informational), Oct. 2004. [Online]. Available:
  \url{http://www.ietf.org/rfc/rfc3954.txt}
\BIBentrySTDinterwordspacing

\bibitem{zhong2021burstsketch}
Z.~Zhong, S.~Yan, Z.~Li, D.~Tan, T.~Yang, and B.~Cui, ``Burstsketch: Finding
  bursts in data streams,'' in \emph{Proceedings of the 2021 International
  Conference on Management of Data}, 2021, pp. 2375--2383.

\bibitem{caida201810}
\BIBentryALTinterwordspacing
CAIDA, ``{The CAIDA UCSD Anonymized Internet Traces - Oct. 18th.}'' 2018.
  [Online]. Available:
  \url{http://www.caida.org/data/passive/passive\_dataset.xml}
\BIBentrySTDinterwordspacing

\bibitem{chen2006collaborative}
Y.~Chen and K.~Hwang, ``Collaborative detection and filtering of shrew {DDoS}
  attacks using spectral analysis,'' \emph{Journal of Parallel and Distributed
  Computing}, vol.~66, no.~9, pp. 1137--1151, 2006.

\bibitem{fu2021realtime}
C.~Fu, Q.~Li, M.~Shen, and K.~Xu, ``Realtime robust malicious traffic detection
  via frequency domain analysis,'' in \emph{Proceedings of the 2021 ACM SIGSAC
  Conference on Computer and Communications Security}, 2021, pp. 3431--3446.

\bibitem{luo2005new}
X.~Luo, R.~K. Chang \emph{et~al.}, ``On a new class of pulsing
  denial-of-service attacks and the defense.'' in \emph{NDSS}, 2005.

\bibitem{chang2010taming}
C.-W. Chang, S.~Lee, B.~Lin, and J.~Wang, ``The taming of the shrew: Mitigating
  low-rate tcp-targeted attack,'' \emph{IEEE Transactions on Network and
  Service Management}, vol.~7, no.~1, pp. 1--13, 2010.

\bibitem{Estan03}
C.~Estan and G.~Varghese, ``New directions in traffic measurement and
  accounting: Focusing on the elephants, ignoring the mice,'' \emph{ACM
  Transactions on Computer Systems}, vol.~21, no.~3, pp. 270--313, 2003.

\bibitem{sivaraman2017heavy}
V.~Sivaraman, S.~Narayana, O.~Rottenstreich, S.~Muthukrishnan, and J.~Rexford,
  ``Heavy-hitter detection entirely in the data plane,'' in \emph{Proceedings
  of the Symposium on SDN Research}.\hskip 1em plus 0.5em minus 0.4em\relax
  ACM, 2017, pp. 164--176.

\bibitem{scherrer2021lowrate}
S.~Scherrer, C.-Y. Wu, Y.-H. Chiang, B.~Rothenberger, D.~E. Asoni, A.~Sateesan,
  J.~Vliegen, N.~Mentens, H.-C. Hsiao, and A.~Perrig, ``{Low-Rate Overuse Flow
  Tracer (LOFT): An Efficient and Scalable Algorithm for Detecting Overuse
  Flows},'' in \emph{Proceedings of the Symposium on Reliable Distributed
  Systems (SRDS)}, 2021.

\bibitem{huang2021toward}
Q.~Huang, S.~Sheng, X.~Chen, Y.~Bao, R.~Zhang, Y.~Xu, and G.~Zhang, ``Toward
  nearly-zero-error sketching via compressive sensing,'' in \emph{18th {USENIX}
  Symposium on Networked Systems Design and Implementation ({NSDI} 21)}, 2021,
  pp. 1027--1044.

\bibitem{sheng2021pr}
S.~Sheng, Q.~Huang, S.~Wang, and Y.~Bao, ``{PR-Sketch: monitoring per-key
  aggregation of streaming data with nearly full accuracy},'' \emph{Proceedings
  of the VLDB Endowment}, vol.~14, no.~10, pp. 1783--1796, 2021.

\bibitem{yang2019heavykeeper}
T.~Yang, H.~Zhang, J.~Li, J.~Gong, S.~Uhlig, S.~Chen, and X.~Li,
  ``{HeavyKeeper: An Accurate Algorithm for Finding Top-$ k $ Elephant
  Flows},'' \emph{IEEE/ACM Transactions on Networking}, vol.~27, no.~5, pp.
  1845--1858, 2019.

\bibitem{liu2016one}
Z.~Liu, A.~Manousis, G.~Vorsanger, V.~Sekar, and V.~Braverman, ``One sketch to
  rule them all: Rethinking network flow monitoring with {UnivMon},'' in
  \emph{Proceedings of the 2016 ACM SIGCOMM Conference}, 2016, pp. 101--114.

\bibitem{niestegge1990leaky}
G.~Niestegge, ``The ‘leaky bucket’policing method in the atm (asynchronous
  transfer mode) network,'' \emph{International Journal of Digital \& Analog
  Communication Systems}, vol.~3, no.~2, pp. 187--197, 1990.

\bibitem{Wu14}
H.~Wu, H.-C. Hsiao, and Y.-C. Hu, ``Efficient large flow detection over
  arbitrary windows: An algorithm exact outside an ambiguity region,'' in
  \emph{Proceedings of the 2014 Conference on Internet Measurement Conference
  (IMC)}.\hskip 1em plus 0.5em minus 0.4em\relax ACM, 2014, pp. 209--222.

\bibitem{sateesan2020novel}
A.~Sateesan, J.~Vliegen, J.~Daemen, and N.~Mentens, ``Novel {Bloom} filter
  algorithms and architectures for ultra-high-speed network security
  applications,'' in \emph{2020 23rd Euromicro Conference on Digital System
  Design (DSD)}.\hskip 1em plus 0.5em minus 0.4em\relax IEEE, 2020, pp.
  262--269.

\bibitem{daemen2018xoodoo}
J.~Daemen, S.~Hoffert, G.~Van~Assche, and R.~Van~Keer, ``Xoodoo cookbook.''
  \emph{IACR Cryptol. ePrint Arch.}, vol. 2018, p. 767, 2018.

\bibitem{xilinx2021}
Xilinx, ``Xilinx {V}irtex {U}ltra{S}cale+ {FPGA} {VCU118} {E}valuation {K}it,''
  \url{https://www.xilinx.com/products/boards-and-kits/vcu118.html}, 2021.

\bibitem{popoviciu2007}
C.~P. Popoviciu, E.~Levy-Abegnoli, and P.~Grossetete, ``{N}etwork {P}erformance
  {C}onsiderations: {C}oexistence of {IP}v4 and {IP}v6,''
  \url{https://bit.ly/3BHWIIc}, 2007.

\bibitem{bosshart2014p4}
P.~Bosshart, D.~Daly, G.~Gibb, M.~Izzard, N.~McKeown, J.~Rexford,
  C.~Schlesinger, D.~Talayco, A.~Vahdat, G.~Varghese \emph{et~al.}, ``P4:
  Programming protocol-independent packet processors,'' \emph{ACM SIGCOMM
  Computer Communication Review}, vol.~44, no.~3, pp. 87--95, 2014.

\bibitem{albusp4code}
\BIBentryALTinterwordspacing
S.~Scherrer, ``{ALBUS} {P}4 {C}ode,'' 2023. [Online]. Available:
  \url{https://github.com/simonschdev/srds23-albus-p4}
\BIBentrySTDinterwordspacing

\bibitem{v1model}
\BIBentryALTinterwordspacing
{Barefoot Networks Inc.}, ``{V1M}odel {S}witch,'' 2021. [Online]. Available:
  \url{https://github.com/p4lang/p4c/blob/main/p4include/v1model.p4}
\BIBentrySTDinterwordspacing

\bibitem{Liu2016}
Z.~Liu, A.~Manousis, G.~Vorsanger, V.~Sekar, and V.~Braverman, ``{One Sketch to
  Rule Them All: Rethinking Network Flow Monitoring with UnivMon},'' in
  \emph{ACM SIGCOMM}, 2016.

\bibitem{yang2018elastic}
T.~Yang, J.~Jiang, P.~Liu, Q.~Huang, J.~Gong, Y.~Zhou, R.~Miao, X.~Li, and
  S.~Uhlig, ``Elastic sketch: Adaptive and fast network-wide measurements,'' in
  \emph{Proceedings of the 2018 Conference of the ACM Special Interest Group on
  Data Communication}.\hskip 1em plus 0.5em minus 0.4em\relax ACM, 2018, pp.
  561--575.

\bibitem{boyer1991mjrty}
R.~S. Boyer and J.~S. Moore, ``Mjrty—a fast majority vote algorithm,'' in
  \emph{Automated Reasoning}.\hskip 1em plus 0.5em minus 0.4em\relax Springer,
  1991, pp. 105--117.

\bibitem{rao2011network}
A.~Rao, A.~Legout, Y.-s. Lim, D.~Towsley, C.~Barakat, and W.~Dabbous, ``Network
  characteristics of video streaming traffic,'' in \emph{Proceedings of the
  seventh conference on emerging networking experiments and technologies},
  2011, pp. 1--12.

\end{thebibliography}

\clearpage

\ifthenelse{\boolean{fullpaper}}{
    \appendix
    \subsection{Accuracy Analysis}
\label{sec:analysis:accuracy}

The detection accuracy of \name{} is measured
by its ability to avoid false positives
and false negatives. Since the LB
algorithm is perfectly accurate in that sense,
all inaccuracy stems
from the dynamic selection of the LB-monitored flows.

\textbf{False positives.} \name{} preserves the zero false positives
of the LB algorithm. 
By definition from~\cref{sec:problem:definition}, 
a flow violates a flow specification~$\gamma t + \beta$ 
if there exists a time window with length~$w$ in which
the flow volume exceeds~$\gamma w+\beta$.
If~\name{} reports a flow,
this flow was monitored during a time window with length~$w'$
and had a volume of more than~$\gamma w' + \beta$.
Hence, since~\name{} reports a flow only if it
encountered a time window in which the volume allowance is exceeded,
\name{} never reports flows which do not violate the flow specification,
i.e., \name{} has zero false positives.

\textbf{False negatives.}
To formally characterize the occurrence of
false negatives in~\name{}, we consider a burst of a flow~$f$
with width~$w$, overuse ratio~$\ell$, and
shape~$v_f(t)$, a fluid approximation of the volume sent
until time~$t \in [0, w]$ after the burst start.
Hence, it holds that~$v_f(0) = 0$ and~$v_f(w) = \gamma w+\ell\beta$.
Moreover, we consider bursts with~$v_f'(t) := d/dt\ v_f(t) \geq \gamma\ \forall t \in [0,w]$,
as any burst can be decomposed into shorter bursts for which this
property holds. As a result, it  holds that~$v(t) \geq \gamma t$
for all~$t \in [0, w]$.
If such a burst is assigned to an LB in~\name{} at time~$t'$,
\name{} observes the burst volume~$v_f(w) - v_f(t')$,
and reports the flow if the observed volume exceeds~$\gamma (w-t')+\beta$.
Hence, \name{} detects the excessively bursty flow
if the flow is assigned to an LB at time~$t'$, where
\begin{equation}
    v_f(t') < \gamma t' + (\ell-1)\beta.
    \label{eq:det-condition}
\end{equation}

Since~$v'_f(t) > \gamma$ for all~$t$,
there exists a unique time~$t^* \in (0,w)$ such that
for all~$t' \in [0, t^*)$, the detection condition
in~\cref{eq:det-condition} holds.
Hence,~$t^*$ denotes the time
from which detection is not achieved anymore.
\name{} provides two options how a
flow might be assigned to an LB before time~$t^*$.

The first possibility for assigning the 
excessively bursty flow~$f$ to LB~$\lambda$
is \emph{push-based}: If a time~$t'$ exists
such that~$f$ is assigned to the BF~$\pi$ and the
push threshold is exceeded by the BC count~$\pi.c(t') > T$,
flow~$f$ is inserted into LB~$\lambda$.
With the rigidity~$r = 0$ used in our experiments,
flow~$f$ can only reach such a high BC count
if it outweighs the aggregate traffic volume
of all other flows~$g \in F_{\pi}$, $g\neq f$, mapped to~$\pi$.
Let~$v_g(t')$ be the volume sent by flow~$g$ 
between the burst start of flow~$f$ (i.e., $t=0$) and time~$t'$.
Importantly, the BC count~$\pi.c(0)$ of
the flow~$g' := \pi.f(0)$ is subsumed into~$v_{g'}$.
In formal terms, the detection probability~$p(f)$
of flow~$f$ is 1 (i.e., detection is guaranteed)
under the following condition:
\begin{equation}
\begin{split}
    p(f) = 1 \iff &\exists t' \in [0, t^*).\ v_f(t') > T + \sum_{\substack{g \in F_{\pi}.\\g\neq f}} v_g(t')\\
    &=: P_{\mathit{push}}(f)
\end{split}
\end{equation}

Second, even if a push is not possible, detection
may still be possible via the \emph{pull-based} approach,
i.e., if the LB~$\lambda$ is cleared
at a time~$t' \in [0, t^*)$ and flow~$f$ occupies BC~$\pi$ at 
that time, i.e., $\pi.f(t') = f$.
To characterize the probability of~$\pi.f(t') = f$, 
we need to consider all flows~$g \in F_{\pi}$ mapped to BC~$\pi$.
Given such flow volumes~$\{v_g(t)\}_{g\in F_{\pi}}$
for any~$t$, we note that the probabilistic-decay technique
is equivalent to the Majority algorithm~\cite{boyer1991mjrty}
for rigidity~$r = 0$, which we use in our experiments.
This algorithm is guaranteed to find the majority item
in arbitrary streams, if such an element exists. If no
majority item exists, the algorithm outputs item~$f$
with a probability that corresponds to the volume
of item~$f$ compared to the volume of all other items
in the stream. 
Applied to our setting, the probability
that the bursty flow~$f$ occupies BC~$\pi$ at time~$t'$
is given as follows:
\begin{align}
    \mathrm{P}\left[\pi.f(t') = f\right] &= p_{M}(v_f(t'), t') ,
    \label{eq:analysis:BC-probability}\\
     \text{where } p_{M}(v, t') &= \min\left(1,\ \frac{ v }{\sum_{g\in F_{\pi}.\ g\neq f} v_g(t')}\right).
    \label{eq:analysis:majority-algorithm-probability}
\end{align}
Notably, multiple LB-clearing moments
may provide opportunity for a pull, i.e.,
the occupancy probability above is relevant
at all moments $\{t_i'\}_{i \in \mathbb{N}, i \leq \tau}$, $t_i' \in [0, t^*)$, $\tau \geq 0$.
The cumulative probability of LB pulls with
subsequent detection is thus:
\begin{equation}
    p_{\mathit{pull}}(f) = \sum_{\substack{i \in \mathbb{N}\\ i \leq \tau}} p_M(v_f(t_i'), t_i') \prod_{\substack{j \in \mathbb{N}\\j < i}} (1 - p_M(v_f(t_j'), t_j'))
\end{equation}

In summary, the detection probability~$p(f)$ is given by:
\begin{equation}
    p(f) = \begin{cases}
        1 & \text{if }  P_{\mathit{push}}(f),\\
        p_{\mathit{pull}}(f) & \text{otherwise,}
    \end{cases}
\end{equation}
which is only zero if~$\neg P_{\mathit{push}}(f)$
and~$\tau = 0$ (i.e., the LB~$\lambda$ is never cleared).
Therefore, even if the adversary manages to prevent LB clearings
with a masking attack, the attack damage in this scenario
is bounded because~$P_{\mathit{push}}$ becomes true
for high enough attack rates.
    \subsection{Evaluation of Background Filtering}
\label{sec:evaluation:prefiltering}

\begin{figure}
    \centering
    \captionsetup{justification=centering}
    \begin{subfigure}{0.38\linewidth}
        \centering
        \includegraphics[width=\linewidth,trim=0 5 0 0]{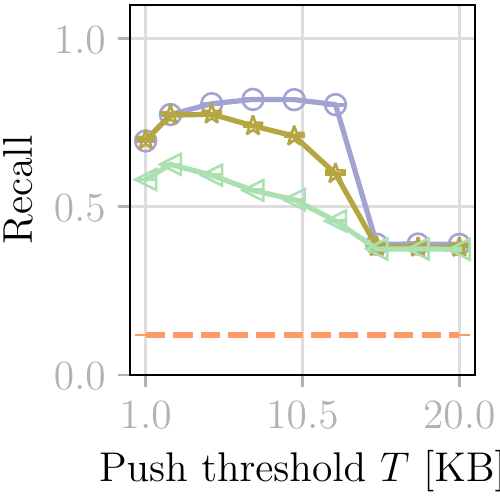}
        \caption{200ms bursts.}
        \label{fig:evaluation:prefiltering:200}
    \end{subfigure}
    \begin{subfigure}{0.38\linewidth}
        \centering
        \includegraphics[width=\linewidth,trim=0 5 0 0]{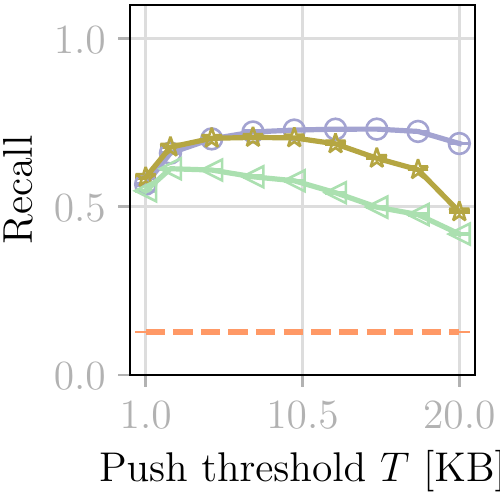}
        \caption{400ms bursts.}
        \label{fig:evaluation:prefiltering:400}
    \end{subfigure}\\
    \begin{subfigure}{\linewidth}
        \centering
        \includegraphics[width=\linewidth,trim=170 50 200 620,clip]{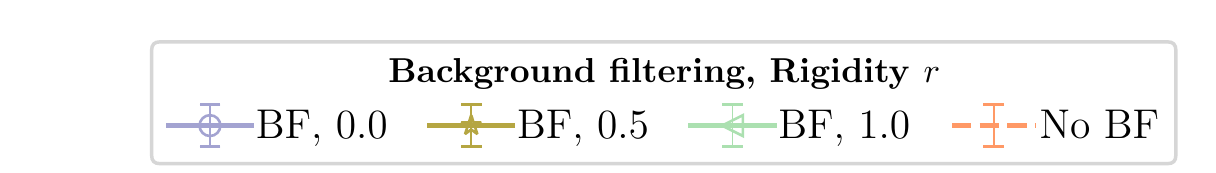}
    \end{subfigure}
    \caption{Background-filtering evaluation.}
    \label{fig:evaluation:prefiltering}
\end{figure}

In this section, we test the background filter
on its effects on detector accuracy.
Apart from its safeguard function, the background filter
also aims at singling out bursty flows that are worth monitoring 
for the leaky buckets.
To test whether the background filter lives up to this expectation, 
we repeat the base-configuration experiment from~\cref{sec:evaluation:sensitivity},
but vary the background0filtering configuration.
In particular, we evaluate \name{} with and without background filtering, 
with varying rigidity~$r$ and with varying push threshold~$T$. 
For the version without background filtering, 
the memory formerly allocated to background counters is
instead allocated to additional leaky buckets
to hold the total memory allocation constant.
The recall results of the evaluation are shown in~\cref{fig:evaluation:prefiltering}
(Perfect precision is consistently achieved by design).

\textbf{Background filtering in general.}
We observe that background filtering boosts recall considerably.
All investigated background-filtering configurations achieve
substantially higher recall than \name{}
without background filtering. Hence, the 
background filter is a vital part of~\name{}.

\textbf{Push threshold.} The results confirm the expected trade-off:
If the push threshold is too high, recall decreases 
because bursty flows fail to evict the current 
LB-monitored flow.
Conversely, if the push threshold is too low, 
flows evict each other from the LB
before any bursty flow is detected, 
also harming recall.
Moreover, both the optimal~$T$ and the
recall considerably depend on the rigidity~$r$,
where we have found the minimum rigidity~$r=0$
to be optimal.
Given the varying burst width in~\cref{fig:evaluation:prefiltering:200,fig:evaluation:prefiltering:400},
we observe that the push threshold should be set low if
short bursts are to be detected, e.g., not above 12~KB
if bursts of duration 200~ms should be detected. 
However, we also observe that a low push threshold
does not strongly compromise the ability of~\name{}
to detect longer bursts. For example, if the 
push threshold is set to 10~KB to optimize recall of 200~ms bursts,
the recall of 400~ms bursts is still near-optimal.
    \subsection{Memory Analysis}
\label{sec:memory-analysis}

The following analysis justifies the claim
of 16 bytes per LB-BC pair in~\cref{sec:analysis:complexity:memory}.

\textbf{LB/BC flow ID.} Empirical data suggests a concurrency
of around 10 million flows for 1 Tbps of forwarding 
capacity~\cite{caida201810}, which implies that
a field size of 3 bytes is sufficient to distinguish the expected number of flows.

\textbf{LB timestamp.}
4 bytes are sufficient to accommodate
nanosecond timestamps with a timestamp-reset period
of around 4.3 seconds. 
Given two timestamps in known order,
the correct difference between the two timestamps
can be calculated if these timestamps
differ by less than the timestamp-reset period,
even if an overflow takes place between
the two timestamps. 
In a realistic environment,
flows that send no packet for more than 4.3
seconds are not considered bursty and are
evicted by the time-out safeguard
before this maximum difference is exceeded.

\textbf{LB count.}
The LB count field
must be able to record the maximum burstiness
volume~$\beta$. If operating with
a rate allowance of~$\gamma = 1\text{ Mbps}$
and tolerating flows that send at most~5 
times rate~$\gamma$ during 100 milliseconds, 
the burstiness allowance~$\beta$ corresponds
to 50 kilobytes, which can in turn be counted
by fields of 2 bytes. 

\textbf{BC count.} The BC count field
must be able to record probabilistically
decaying flow volume
up to the push threshold~$T$.
This threshold is maximal in the (artificial) 
case~$r = \infty$, where
the BC count is never decremented.
The threshold~$T$ in this maximum case
limits the acceptable
flow volume during the time~$w'$ 
in which another flow is in the LB.
Assuming again that
the algorithm targets flows sending at more than
5~times the allowed rate for $w = 100$ milliseconds, 
and that flows occupy a LB for 
at most 100ms ($w' = w$), a reasonable choice
for the threshold~$T$ 
in the maximum case~$r = \infty$ 
is $5\gamma w = 62.5$ kilobytes,
which can also be counted with a 2-byte field.

In summary, 14 bytes are needed for
an LB-BC combination, which we increase
to 16 bytes for alignment.
This low memory consumption ensures
high memory efficiency: \cref{sec:evaluation:sensitivity}
suggests that a memory consumption of 300KB
(on the order of usual L1/L2 cache sizes)
allows \name{} to outperform previous algorithms
(when considering both recall and precision)
on a 10~Gbps link~\cite{caida201810}.
    \subsection{P4 Implementation Details}
\label{sec:implementation:p4-details}

We created an additional implementation in
P4~\cite{bosshart2014p4}, which allows to embed
complex functionality in programmable 
switches~\cite{agrawal2020intel}.
Our P4 code is available online~\cite{albusp4code} 
and can be tested using the virtual V1Model 
switch~\cite{v1model}.

While \name{} is meant to be used as a monitoring primitive
within sophisticated DDoS defense systems 
(mostly also implemented in P4), 
we have created a stand-alone implementation of the algorithm for testing
(cf.~\cref{fig:implementation:p4}). First, an incoming packet
is matched against the blacklist, which is embodied
in a match-action table. If the blacklist matches the
flow attributes in the packet, the packet is dropped;
otherwise, the \name{} algorithm is applied with
an action containing the \name{} control flow. 
If the packet flow is excessively
bursty, the \name{} update sets a metadata flag
in the packet. If a subsequent match-action table matches
this flag, blacklisting is triggered 
by sending the flow attributes to the controller via a digest operation
(Digestion is not possible within the
\name{} update action). The controller then installs a rule including the
malicious-flow attributes into the blacklist match-action table.

\begin{figure}
    \centering
    \begin{tikzpicture}[
flowsquare/.style={draw=black!60, shading=radial,outer color={rgb,255:red,137;green,207;blue,240},inner color=white, thick, minimum size=\nodewidth,outer sep=0pt},
switchsquare/.style={draw=black!60, shading=radial,outer color={rgb,1:red,0.95;green,0.85;blue,0.75},inner color=white, thick,minimum width=46mm,minimum height=25mm,align=center},
tablesquare/.style={draw=black!60, shading=radial,outer color={rgb,1:red,1.0;green,0.6;blue,0.4},inner color=white, thick,minimum height=6mm,align=center},
actionsquare/.style={draw=black!60, shading=radial,outer color={rgb,1:red,1.0;green,0.8;blue,0.4},inner color=white, thick,minimum height=6mm,align=center},
controllersquare/.style={draw=black!60, shading=radial,outer color={rgb,1:red,0.67;green,0.88;blue,0.69},inner color=white, thick,minimum width=46mm,minimum height=6mm,align=center},
matchsquare/.style={fill,color={rgb,1:red,0;green,0.5;blue,0.0},text=white,},
nomatchsquare/.style={fill,color={rgb,1:red,0.5;green,0;blue,0},text=white,},
]

    \def\nodewidth{6mm};
    \def\counterwidth{6mm};
    \def\indexheight{5mm};
    
    \draw[dashed] (-0.4, 3.1) -- (7.5, 3.1);
    \node[align=right,color={rgb,1:red,0.5;green,0.5;blue,0.5}] at (7.0, 3.5) {Control\\plane};
    \node[align=right,color={rgb,1:red,0.5;green,0.5;blue,0.5}] at (7.1, 2.6) {Data\\plane};
    
    \node[flowsquare] (f1) at (0, 0.8) {$f_1$};
    
    \node[switchsquare] (switch) at (3.45, 1.75) {};
    \node[] (switch_text) at (3.45, 0.8) {Switch};
    
    \node[align=right,color={rgb,1:red,0.5;green,0.5;blue,0.5}] at (0.6, 1.5) {Tables};
    \node[tablesquare] (blacklist) at (2, 1.5) {blacklist};
    \node[tablesquare] (check_block) at (4.6, 1.5) {check\_block};
    
     \node[align=right,color={rgb,1:red,0.5;green,0.5;blue,0.5}] at (0.5, 2.5) {Actions};
    \node[actionsquare] (drop) at (1.75, 2.5) {drop};
    \node[actionsquare] (albus_update) at (3.4, 2.5) {albus\_update};
    \node[actionsquare] (block) at (5.1, 2.5) {block};
    
    \node[controllersquare] (controller) at (3.45, 3.5) {Controller};
    
    \draw[-latex] (f1.east) -- (2.0, 0.8) -- (blacklist.south);
    
    \draw[-latex] (blacklist.57) -- (drop.south east);
    \node[matchsquare] (blacklist_match) at (1.9, 2.0) {\tiny\CheckmarkBold};
    
    \draw[-latex] (blacklist.42) -- (albus_update.south west);
    \node[nomatchsquare] (blacklist_nomatch) at (2.65, 2.0) {\tiny\XSolidBold};
    
    \draw[-latex] (albus_update.298) -- (check_block.north west);
    
    \draw[-latex] (check_block.95) -- (block.south west);
    \node[matchsquare] (check_block_match) at (4.9, 2.0) {\tiny\CheckmarkBold};
    
    \draw[-latex] (check_block.265) -- (4.57, 0.8) -- (6.6, 0.8);
     \node[nomatchsquare] (check_block_nomatch) at (4.9, 1.0) {\tiny\XSolidBold};
    
    \draw[-latex,densely dotted,thick] (block.95) -- (controller.349);
    \node[draw=black!60,fill=white,text=black] at (5.75,3.15) {digest};
    
    \draw[-] (controller.349) -- (5.05,3.3) -- (1.4,3.3) -- (1.4,3.2);
    
    \draw[-latex,densely dashed,thick] (1.4,3.2) -- (blacklist.153);
    \node[draw=black!60,fill=white,text=black] at (0.7,3.15) {install};
    
\end{tikzpicture}
    \caption{P4 implementation design.}
    \vspace{-10pt}
    \label{fig:implementation:p4}
\end{figure}
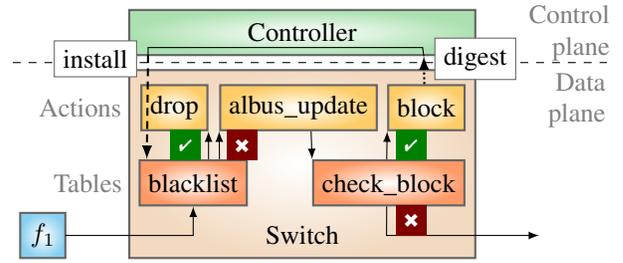

    \subsection{Evaluation on Synthetic Trace}
\label{sec:syn-evaluation}

\begin{figure*}
    \centering
    \begin{minipage}{0.56\linewidth}
    \captionsetup{justification=centering}
    \begin{subfigure}{0.325\linewidth}
        \centering
        \includegraphics[width=\linewidth,trim=0 5 0 0]{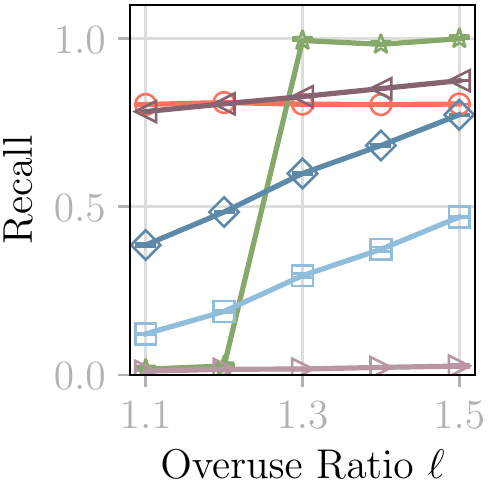}
        \caption{Recall.}
        \label{fig:evaluation:sensitivity:synthetic:recall}
    \end{subfigure}
    \begin{subfigure}{0.325\linewidth}
        \centering
        \includegraphics[width=\linewidth,trim=0 5 0 0]{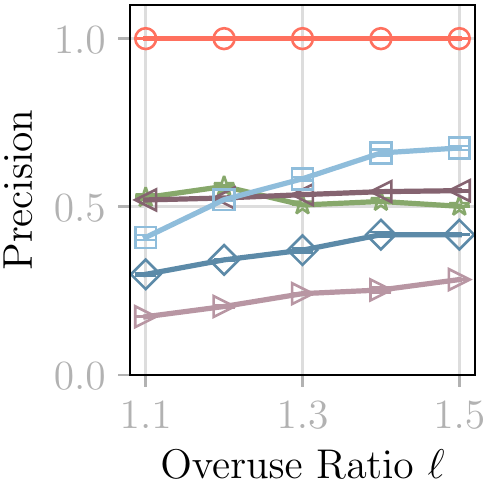}
        \caption{Precision.}
        \label{fig:evaluation:sensitivity:synthetic:precision}
    \end{subfigure}
    \begin{subfigure}{0.325\linewidth}
        \centering
        \includegraphics[width=\linewidth,trim=0 5 0 0]{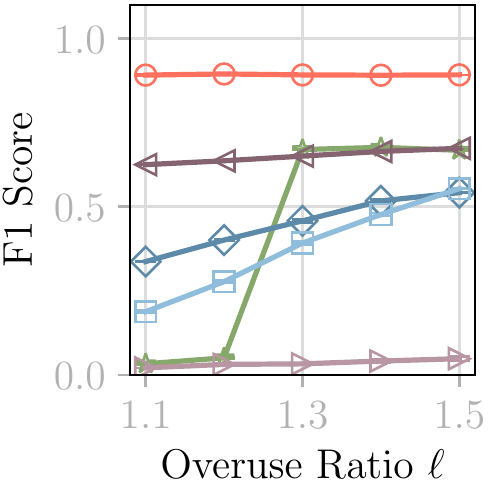}
        \caption{F1 score.}
        \label{fig:evaluation:sensitivity:synthetic:f1}
    \end{subfigure}
    \end{minipage}
    \vspace{-5pt}
    \caption{Evaluation on synthetic background traffic.}
    \label{fig:evaluation:sensitivity:synthetic}
        \begin{subfigure}{\linewidth}
        \centering
        \includegraphics[width=\linewidth,trim=1 50 1 620,clip]{fig/caida_legend.pdf}
    \end{subfigure}
\end{figure*}

In the CAIDA trace used in~\cref{sec:evaluation},
most background flows are short and small, i.e., only consist of 
a few packets within one second. To evaluate a different scenario,
we synthesize background traffic with relatively large and long-lived
background flows, i.e., each flow in the background traffic sends 
exactly at the allowed rate~$\gamma = 1$~Mbps during the whole experiment.
Such traffic can be present on links delivering
a high amount of streamed video~\cite{rao2011network}.
Using this background traffic, we again vary the overuse ratio
of the bursts in the attack, leading to the results 
in~\cref{fig:evaluation:sensitivity:synthetic}.
While the recall results are largely similar to~\cref{fig:evaluation:sensitivity:or:recall}
(same experiment with CAIDA traffic), 
BurstSketch has perfect recall on
high overuse ratios, again because the long-lived background flows 
increase contention in the first stage of BurstSketch. 
\name{} achieves a slightly lower recall than CountMin-Sketch(0.5)
on high overuse ratios, but substantially outperforms
it in both precision and F1 score, an aggregation
of recall and precision (\cref{fig:evaluation:sensitivity:synthetic:f1}).
}{
    
}

\end{document}